\documentclass[Normal]{svjour3}                     
\smartqed  
\pdfoutput=1
\usepackage{graphicx}
\usepackage{amsmath}
\usepackage{hyperref}
\usepackage{tabularx} 
\usepackage{booktabs} 
\usepackage{hhline}
\usepackage{array}
\newcolumntype{L}[1]{>{\raggedright\let\newline\\\arraybackslash\hspace{0pt}}m{#1}}
\usepackage{subfigure}
\usepackage{enumitem}
\usepackage{multirow}
\usepackage{comment}
\usepackage{multirow}

\usepackage{algpseudocode}
\usepackage{algorithm} 
\usepackage{pgfplots}

\usepackage{lscape} 
\usepackage{amssymb}
\usepackage{cite}

\makeatother
\definecolor{armygreen}{rgb}{0.0, 0.5, 0.13}

\journalname{Empirical Software Engineering}

\newcolumntype{C}[1]{>{\centering\let\newline\\\arraybackslash\hspace{0pt}}m{#1}}
\newcolumntype{R}[1]{>{\hfill\let\newline\\\arraybackslash\hspace{0pt}}m{#1}}

\algnewcommand{\LineComment}[1]{\State \(\triangleright\) #1}

\newcommand{\companyname}{SRBD}

\begin{document}

\title{Using a Balanced Scorecard to Identify Opportunities to Improve Code Review Effectiveness: An Industrial Experience Report}

\titlerunning{Improving Effectiveness of Code Reviews}        

\author{Masum Hasan \and
    Anindya Iqbal \and
        Mohammad Rafid Ul Islam \and 
        A.J.M. Imtiajur Rahman \and
        Amiangshu Bosu
}

\institute{M. Hasan, M. Islam, and A. Iqbal \at
              Department of Computer Science and Engineering, \\ Bangladesh University of Engineering and Technology, Dhaka Bangladesh \\
              \email{masum@ra.cse.buet.ac.bd, rafid@openrefactory.com, anindya@cse.buet.ac.bd}           
           \and
           A.J.M. Rahman \at
              Samsung R\&D Institute Bangladesh, Dhaka, Bangladesh\\
              \email{m.imtiaz@samsung.com}
              \and
              Amiangshu Bosu \at
              Department of Computer Science,\\ Wayne State University, Detroit, Michigan, USA\\
              \email{amiangshu.bosu@wayne.edu}
}

\date{Received: January 16, 2021 / August 12, 2021}

\maketitle

\begin{abstract}
Peer code review is a widely adopted software engineering practice to ensure code quality and ensure software reliability in both the commercial and open-source software projects. Due to the large effort overhead associated with practicing code reviews, project managers often wonder, if their code reviews are effective and if there are improvement opportunities in that respect. Since project managers at Samsung Research Bangladesh (SRBD) were also intrigued by these questions, this research  developed, deployed, and evaluated a production-ready solution using the \textit{Balanced SCorecard (BSC) strategy } that SRBD managers can use in their day-to-day management to monitor individual developer's, a particular  project's or the entire organization's code review effectiveness.
Following the four-step framework of the BSC strategy, we-- 1) defined the operation goals of this research, 2) defined a set of metrics to measure the effectiveness of code reviews, 3) developed an automated mechanism to measure those metrics, and 4) developed and evaluated a monitoring application to inform the key stakeholders. 

Our automated model to identify useful code reviews achieves 7.88\% and 14.39\% improvement  in terms of accuracy and minority class $F_1$ score respectively over the models proposed in prior studies. It also outperforms human evaluators from SRBD, that the model replaces, by a margin of 25.32\% and 23.84\% respectively in terms of accuracy and minority class $F_1$ score. 
In our post-deployment survey, SRBD developers and managers indicated that they found our solution as useful and it provided them with important insights to help their decision makings.



\end{abstract}

\section{Introduction}
\label{sec:introduction}

Peer Code Review is a software development practice that enables developers to systematically inspect peers' code changes to  identify potential defects as well as quality improvement opportunities~\cite{bacchelli2013expectations}.
Software development projects, both Open Source Software (OSS) and commercial, are spending a significant amount of resources in code reviews, as developers spend 10-15\% of their time preparing code for reviews or reviewing others code~\cite{Bosu-et-al-TSE}.  Besides finding defects, known benefits of  code reviews include: improving project maintainability~\cite{Bosu-et-al-TSE}, maintaining code integrity~\cite{cohen2006best}, improving the relationships between the participants~\cite{Bosu-et-al-TSE}, and spreading knowledge, expertise, and development techniques among the review participants~\cite{bacchelli2013expectations,rigby2013convergent}.
Despite its widespread usage and numerous benefits, a study at Microsoft found that nearly one-third of the code reviews were considered as `not useful' by the code authors~\cite{bosu}. Due to the large effort overhead associated with practicing code reviews, project managers often wonder, \emph{if their code reviews are effective and if there are improvement opportunities in that respect}~\cite{bosu,czerwonka2015code}. 

These very questions also intrigued the managers at Samsung Research and Development Institute Bangladesh (also known as Samsung Research Bangladesh or SRBD).
To motivate useful code reviews, SRBD introduced a bimonthly \textit{Best Reviewer Award}. To identify the best reviewer(s), multiple members from the Code Quality Assessment (CQA) team spent a significant amount of time manually evaluating the usefulness of all review comments authored during the last two months. However, this process was not only time consuming but also was inconsistent due to the lack of a formal evaluation guideline. Therefore, SRBD managers preferred: 1) an automated and reliable process to identify the best reviewer(s) as well as useful code reviews, 2)  an automated framework to measure the code review effectiveness of a developer, a project, or the entire organization, 3) a mechanism to identify opportunities to improve code review effectiveness. 
To fulfill these three objectives, we proposed and evaluated a solution based on  the \textit{\underline{B}alanced \underline{SC}orecard  (BSC) strategy}, which was conceived by Kaplan and Norton~\cite{balancedscorecard} and has been adopted by more than half of the large firms in the US~\cite{marr2003automating}. 
Prior studies have also reported successful deployments of BSC based solutions among software development organisations~\cite{hofner2011fostering,mair2002balanced,papalexandris2004implementing}. 

This experience report details the development, and evaluation of our solution named  Code Review Analytics (CRA), an internal web application, that SRBD managers use regularly to monitor  code review effectiveness and to identify potential improvement opportunities.
Although the ultimate objective of CRA is continuous improvement of code review effectiveness, it does not address that directly. CRA aims to provide an automated mechanism to access measures and reports to identify improvement opportunities to help adoption of new initiatives. Whether a new initiative can address a problem area and improve code review effectiveness, depends on multiple factors including, the characteristics of the problem identified and the effectiveness of proposed initiatives. 

Our post-deployment evaluation suggests that in addition to fulfilling  the three primary objectives of this research,  the CRA tool is helping SRBD developers : \emph{to improve as reviewers}, \emph{ in measuring the progress of the organization in terms of achieving more effective reviews},  and \emph{in improving the organization culture on code reviews}.
While CRA has  helped SRBD managers in identifying multiple improvement opportunities and  motivated multiple initiatives, we are unable to report those due to our non-disclosure agreement (NDA) with SRBD.

\textbf{Contributions:}  The primary contributions of this research include:
\begin{itemize}
\item An empirical study to identify SRBD developers' perceptions of useful code reviews.
\item A set of metrics to measure a reviewer's code review performance.
\item An automated model to identify the usefulness of a code review comment.
\item A replication of a prior software engineering study~\cite{bosu} in a different context.
\item A successful deployment  of the BSC framework in a software development organization for the purpose of improving code review effectiveness.
\item Code review analytics, a dashboard to monitor the effectiveness of code reviews at SRBD.
\item We have made the Jupyter notebook to train and evaluate our automated model to identify useful code reviews publicly available on Github\footnote{We are unable to make the dataset publicly available due to the restrictions imposed by our NDA with SRBD. } at: \url{https://github.com/WSU-SEAL/CRA-usefulness-model}

\end{itemize}

\textbf{Organization:} The  Remainder of this paper is organized as follows.
Section~\ref{sec:background} provides a brief background on peer code review, and an overview of related prior works.
Section~\ref{sec:research-overview} provides an overview of our four step research methodology.
Section~\ref{sec:goal-definition}  defines the goal of this study based on the BSC framework.
Section~\ref{sec:goal-metrics}  identifies a set of metrics to measure the progress to achieve our defined goals.
Section~\ref{sec:measure} describes the development of tools to measure the identified metrics.
Section~\ref{sec:revise} describes the development and evaluation of a BSC dashboard to enable progress monitoring.
Section~\ref{implication} discusses the implications of this study.
Section~\ref{sec:threats} addresses the threats to validity for this research.
Finally, Section~\ref{sec:conclusion} concludes this paper.

\section{Background}
\label{sec:background}

This section provides a brief overview of contemporary tool-based peer code reviews, code review process at SRBD, code review usefulness, and  related prior works.

\subsection{Tool-based Peer Code Review}

Code review is a software development practice to identify defects as well as provide improved suggestions to code written by peers in a team-based software development project. Modern code reviews are more lightweight and informal than traditional Fagan-inspections \cite{Fagan}  and are widely adopted among both commercial and open source software projects \cite{bacchelli2013expectations,bosu,sadowski2018modern,Kononenko-2016}. There are many popular lightweight applications that facilitate tool-based code reviews, such as, Gerrit, Fabricator,  GitHub pull requests, Git Lab pull requests, CodeFlow, ReviewBoard and Crucible. However, the primary workflow of code reviews across all platforms is very similar. When a code author submits a new patch, a reviewer is assigned either manually or using an automated recommendation system. All code review tools provide  some interfaces to view source, and compare multiple patches of a file side by side before submitting a comment.  After evaluation, a reviewer may provide suggestions in the form of  review comments. The code author submits subsequent patches addressing the issues raised in those comments. The code is merged into the main repository when all major issues are resolved and reviewer(s) approve the change.



\subsection{Code Review Usefulness}
Useful code reviews are the ones that help the code authors either explicitly by identifying defects or shortcomings in a code, or implicitly by  disseminating knowledge or provoking a discussion. The usefulness of code reviews in a project can be an indicator of the effectiveness of the code review process in that project \cite{bosu}. A useful code review helps the code under discussion, or helps the team in the long term. `Not useful' code reviews, on the contrary, fails to achieve any of these goals. Although in some occasions the usefulness of a code review can be subjective, for example a review that is useful to a beginner programmer may not be useful to a senior programmer, previous studies suggest that most `useful' and `not useful' reviews follow some patterns which can be identified automatically \cite{Bosu-et-al-TSE,bosu,Kononenko-2016,rahman}.

\subsection{Related Work}
\label{sec:related-work}
In recent years, there has been an influx of studies investigating various aspects of tool-based modern code reviews. In the following, we briefly discuss  some  of those studies that aimed at improving the effectiveness of the code review process. 

In one of the earlier studies on code review, Rigby and German \cite{rigby2006preliminary} proposed a set of code-review metrics (i.e. acceptance rate, reviewer characteristics, top reviewer vs. top committer, review frequency, number of reviewers per patch, and patch size) to characterize code review practices.

In an empirical study, Kononenko et al. \cite{Kononenko-emp} quantitatively investigated the relationships between the reviewers’ code inspections and a set of personal and social factors that might affect the review quality. 
They  found that the quality of the code review process is associated with personal metrics, such as, reviewer workload and experience, and participation metrics, such as, number of developers involved in the project. Same group of authors later conducted an empirical study at Mozilla and analyzed developer's perception of good reviews. They found that a code review is perceived useful based on thoroughness of the feedback, the reviewer’s familiarity with the code, and the perceived quality of the code itself \cite{Kononenko-2016}. In a study involving OSS and Microsoft developers, Bosu et al. \cite{Bosu-et-al-TSE} characterized the human factors in code review (i.e. author's reputation, relationship between an author and a reviewer, etc.) to guide project managers' decisions about the usefulness of code review and help improve code review effectiveness.

Some other studies have focused on characterizing the reviewers in terms of the quality of their code reviews. Hatton \cite{hatton2008testing} found that defect identification capabilities vary significantly between reviewers and the best reviewer is up to 10 times more effective at finding defects than the worst reviewer. He also found that a two person team identified 76\% of the defects on average, where individuals on average identified 53\% defects. Bosu et al. \cite{bosu} showed that reviewer’s prior experience in changing or reviewing the artifact and the reviewer’s project experience increases the likelihood that s/he will provide useful feedback. Ebert et al. \cite{8668024} found that missing rationale, discussion of non-functional requirements of the solution, and lack of familiarity with existing code are some reasons for decrease in review quality.Other factors that impact code review quality include co-working frequency of a reviewer with the patch author ~\cite{thongtanunam-dynamics}, description length of a patch~\cite{thongtanunam2017review}, and the level of agreement among the reviewers~\cite{hirao2016impact}.
Chouchen et al. \cite{9425884} cataloged common poor code review practices and explore and characterize their symptoms, causes, and impacts. Thongtanunam et al. \cite{thongtanunam2015should}, Rahman et al. \cite{rahman-correct}, Fracz et al. \cite{developersgame} and Barnett et al. \cite{barnett2015helping}, Chouchen et al. \cite{whoreview} worked on automatic reviewer recommendation based on code review so that the code review practice is more effective.

While many previous studies have analyzed characteristics of code reviews and reviewers, automatic classification of code review effectiveness has received relatively less attention from researchers. The first study related to code review usefulness classification was performed by Bosu et al. \cite{bosu} in a study involving Microsoft developers. They found that 34.5\% code reviews at Microsoft were perceived as not useful by the code authors. The authors also identified a number of quantitative features (i.e. change trigger, number of participants in a comment thread, reply from the author, number of patches etc.) that contributes to code review quality. Rahman et al. \cite{rahman} created an automatic tool for code review classification using textual features, (i.e. reading ease, stop word ratio, question ratio, code element ratio, conceptual similarity between code and comment, etc.) and developer experience (i.e. code authorship, code reviewership etc.).

While most of these studies discuss the qualitative and quantitative analysis of modern code review practices at OSS and commercial projects, to the best of our knowledge, ours is the first study on industrial application of an automatic code review usefulness analysis tool, challenges regarding building such a tool, its effectiveness and impact.

\section{Research Method}
\label{sec:research-overview}
Since our research framework follows the Balanced SCorecard  (BSC) strategy, this section provides a brief overview of the BSC steps.
BSC suggests achieving objectives based on the following four steps~\cite{balancedscorecard}:

\begin{enumerate}
    \item {Define operational goals to fulfill a vision.}
    \item {Define metrics to measure progress.}
    \item {Take initiatives and measure progress towards achieving defined goals.}
    \item {Inform key individuals automatically about scorecard status, and determine why problems occur to revise strategies.}

\end{enumerate}
Based on the four suggested steps of the BSC framework, we developed a four-step  research method to achieve this study's objectives. The research tasks carried out during the four steps of this study are:

\begin{itemize}
    \item \textbf{Step 1:} \emph{Defining operational goals.} We defined the goal  of this research as: `` Automated monitoring of code review effectiveness in SRBD to help identify improvement opportunities in that respect." 
    
    \item \textbf{Step 2:} \emph{Defining metrics to measure code review effectiveness.} In this step, we defined a set of metrics to measure the effectiveness of code reviews in SRBD during a period.
    
    \item \textbf{Step 3:} \emph{Building a framework to measure the code review effectiveness metrics.}  We developed a Machine Learning model to  automatically evaluate the usefulness of each code review comment. Leveraging this model, we automatically compute the set metrics  defined in \emph{Step 2} to evaluate the performance of reviewers and projects for any particular time period.
    
    \item \textbf{Step 4:} \emph{Build a monitoring mechanism to inform the stakeholders.} We created  Code Review Analytics (CRA), an internal web app, that SRBD managers are able to  use regularly to monitor the performance of individual developers or a particular project.  CRA is being used by SRBD managers to identify best reviewers during a particular period as well as monitor the organization's progress towards achieving defined goals. CRA also helps SRBD managers to identify improvement opportunities and adopt targeted initiatives to mitigate performance barriers. 
\end{itemize}

Each of the following four sections focuses on one of the steps of this research in chronological order.
We describe the study methods and findings separately
for each.

\section{Step 1: Defining Operational Goals}
\label{sec:goal-definition}
According to the step 1 of BSC framework, this step defines the operational goals of this research.
At the initiation of this research, we held several meetings with SRBD managers to understand their code review process, their target objectives, and their current problem areas. Following subsections detail SRBD's code review process and current problem areas.

\subsection{Code Reviews at SRBD}
Based on our interviews with SRBD managers and an introductory survey sent to the developers, we identified the following five aspects of code reviews at SRBD.

\begin{itemize}

    \item \textbf{Process:} SRBD has integrated code reviews in its software development pipeline in 2012. As of 2021, Code review is mandatory at SRBD. Each and every change needs at least two approvals before its merging into the main codebase. One of those approvals must come from a senior level developer (i.e., someone with at least two years of experience in that specific domain). Confusions arising during code reviews are often resolved based on  offline discussions between the reviewer and the author.
    
    \item \textbf{Tool:}  While most of the projects used Gerrit, some of the projects use  GitHub pull requests,  GitLab pull requests, and Swarm. 
    
    \item \textbf{Efforts:} SRBD developers spend on average six and half hours per week in code reviews by either reviewing other developers' code or  responding to reviews for their own code.
    
    \item \textbf{Reviewer selection:} Usually, the author of a change invites the reviewers. Accepting a review request is voluntary. Junior members are encouraged to invite a senior member (e.g. the project lead), or a domain expert. While reviewers are more likely to be selected from the same team,  cross-team reviewers are not uncommon.
    
    \item  \textbf{Rewards:}  Reviewing changes submitted by other developers is one of the criteria to evaluate an employee's performance at SRBD.  

\end{itemize}

\subsection{Problem Areas}

 To promote a better code review culture, SRBD managers had introduced a bi-monthly \emph{Best Reviewer Award}. Every two months a review assessment period of one week is scheduled. During this period multiple members from the Code Quality Assessment (CQA) team spent significant amount of time  to manually evaluate the usefulness of all review comments authored during the last two months. This process not only allowed the managers to identify the best reviewer(s) but also measure the overall effectiveness of code reviews in the company during that period. Based on our conversations with the SRBD managers, we identified the following four challenges with their manual process. 

\begin{enumerate}
\item \textbf{Time-consuming and prone to inconsistencies} Since  manual assessments of code reviews require significant efforts from multiple CQA team members, this process is time-consuming and inefficient. For a large organization such as SRBD, manual review assessment process is not scalable. Moreover, there was no prescribed guidelines for the CQA members on how to assess the usefulness of a review comment. Therefore,  such manual assessments are  prone to biases and inconsistencies.

\item \textbf{Ignoring code contexts lead to inaccuracies}
According to Bosu \emph{et} al., ~\cite{bosu}, to evaluate the usefulness of a code review comment, an independent rater needs to comprehend the code context, read the discussions between the code author and the reviewer(s),  and check  whether the suggestions were incorporated by the author in a subsequent patchset. However,  as the CQA members had to manually assess a large number of code reviews within a short period of time, they rarely investigated associated code contexts and relied primarily on the comment thread. As a result, those evaluations were also prone to inaccuracies.

\item \textbf{Delayed assessment of code review quality} 
SRBD managers usually scheduled a manual assessment  period of code reviews once every two months. 
A project's status may change within these two months and the insights gained from these manual assessment may be already obsolete.
Moreover, SRBD managers may need to wait more than two months to understand the impacts of their new initiatives. They wanted  to access status update more frequently.  Having CQA members working daily on the manual assessments could have been a solution.  However, that would be costly and therefore was not feasible.  An automated framework, may solve this problem by providing  faster access to status updates.

\item \textbf{Difficult to identify areas that need attention} 
The manual assessment reports of code reviews  were created using spreadsheets and were primarily used to identify the quality of reviews and the best reviewers. However, it is very time-consuming to dig through these reports to identify areas of concerns. An in-depth analysis to understand the contemporary review culture in the company,  such as effectiveness of cross-project reviews, optimal time to spend in reviews, and the best reviewers for a particular area, is very difficult to compile from those spreadsheets.

\end{enumerate}


To encounter the four above mentioned challenges, the operational goals of this study is to develop an automated review quality assessment system that:

\begin{enumerate}
 \item  aims to eliminate the requirements of manual assessment efforts.
    \item is consistent, highly accurate, and reliable in identifying effective code reviews.
    \item  can assess a large number of review comments within a short period and provide regular and timely status updates to SRBD managers.
   \item enable a developer monitor his/her code review performance.    
    \item enable identifying areas where management intervention  may be helpful and facilitate follow-up monitoring.
    \item facilitates in-depth analysis to identify improvement opportunities.
      \item  motivates reviewers to write better code reviews by making them aware that their reviews are further analyzed and their effort would be acknowledged (Hawthorne Effect \cite{hawthorne}).
   
\end{enumerate}

\section{Step 2: Defining metrics to measure code review effectiveness}
\label{sec:goal-metrics}
\label{subsec:ranking}
Following step 2 of the BSC framework, this section defines a set of metrics to measure progress towards achieving our goals.
To identify the best reviewer(s) for a period SRBD used two metrics. The first metric, which is called \emph{Comment Count (NC)}, is the total number of review comments made by a developer during a period. The second metric, which is called \emph{Comment Usefulness Density (CUD)}, was proposed by Bosu \textit{et} al~\cite{bosu}, and is measured using the proportion of review comments that are useful. Therefore, 

\begin{equation}
\label{eqn-cud}
CUD = \frac{UC}{NC}
\end{equation}
Where $UC$ = Number of useful comments,
$NC$ = Number of comment made by a developer, and $NC\neq0$

To select the best reviewer, two reviewer rankings were generated, one based on the NC and the other based on the CUD.   Using a developer's position in these two rankings, two scores were computed for him/her, which are:
\begin{itemize}
\item A reviewer's  $NC_{score}$  = ($N+1$ - Position in the NC ranking);  if a developer is among the top $N$ in the NC ranking, else it would be 0.
\item Similarly,    a reviewer's $CUD_{score}$  = $N+1$ - Position in the CUD ranking;  if a developer is among the the top $N$ in the CUD ranking, else it would be 0.

\end{itemize}

Here, $N$ is a hyperparameter that can be selected based on the size and priorities of an organization. $N=30$ was being used at SRBD. Finally, a reviewer's aggregated review score was computed based on the following formula:
\begin{equation}
Review\textsubscript{score} = NC\textsubscript{score} + CUD\textsubscript{score}
\end{equation}

Reviewer(s) with the highest $Review_{score}$  would receive the best reviewer award(s) for a period. While this scoring system takes into account both the number of review comments as well as the usefulness of those comments, we noticed two primary deficiencies in this measure. First, more than one third of code changes in SRBD do not receive any comment during reviews as code changes were `acceptable as it is' to the reviewer(s). The  $Review_{score}$ measure fails to account for review efforts spent on such acceptable code changes. Second, since a developer's $Review_{score}$ is based on his/her positions in two rankings, even if a he/she achieves higher NC and CUD during a period than the last period, his/her $Review_{score}$  still may be lower,  if his/her positions in the rankings degrade and vice versa. As a result, a lower /higher $Review_{score}$  does not necessarily indicate degradation/improvement of a developer's review effectiveness. For the same reason, it is not meaningful to run cross-period comparison using the $Review_{score}$ measure.

In this step, our goal was to define a set of metrics that: 1) would  allow us to  measure the code review impact of a developer both in terms of  review quality and  review quantity,  2) would facilitate cross-period, cross-project, and cross-organization comparisons,  and 3) could be used to track degradation/improvement of review effectiveness.   Based on these requirements, we defined an additional set of metrics. We measure the code review quantity using the number of code reviews a developer  has participated in during a period (NR).
To estimate code review quality, we define a new metric named, \emph{Issue Density (ID)},  which aims to identify reviewers who actively participate in reviews by suggesting improvements or identifying issues. Therefore,

\begin{equation}
\label{eqn-dd}
ID = \frac{UC}{NR}
\end{equation}
Where 
$UC$ = Number of useful comments, $NR$ = Number of code reviews, and $NR\neq 0$

To build a ranking of the reviewers based on both review quantity and review quality, we propose two metrics. The first metric, which we call `Review efficiency (RE)', aims to identify the most effective reviewers combining review quantity and quality measures according to the following equation.
 
\begin{equation*}
RE = log_{2}(NR+ 1) \times  (CUD + ID) 
\end{equation*}
\begin{equation}
\label{eqn-ri}
              = log_{2}(NR+1) \times ( \frac{UC}{NC} + \frac{UC}{NR} )
\end{equation}

We use binary logarithmic value of NR+1 to reduce the impact of the number of reviews a developer has participated on the RE measure. Equation~\ref{eqn-ri} rewards a developer for authoring useful comments by increasing both the first and second terms but penalizes him/her for useless comments by decreasing the first term. A developer's participation in additional reviews without making any comments will have mixed effects by increasing the first term but decreasing the second term. We include $log_{2}(NR+1)$ multiplier to account for reviewers who conducted only one or two reviews but authored several useful comments in those reviews ( e.g., Developer E in Table~\ref{tab:metrics_example}), as we want to recognize reviewers who authors useful comments consistently rather than someone who participate rarely.

The second metric, which we call `Review Impact (RI), aims to identify the most impactful reviewer(s) during a period. This measure, which is motivated based on StackOverflow's reputation scoring system\footnote{On StackOverflow, each accepted answer gets 15 points, upvote gets 10 points, and downvote gets -2 points}~\cite{so-reputation},  awards (+10) point for each review a developer participates, (+15) points for each useful comment, and (-2) points for each useless comment.  Although the RI metric penalizes reviewers' for useless comments, these penalties are 7.5 times lower than the rewards from a useful comment and therefore may not dissuade a reviewer from authoring his/her concerns.

\begin{equation*}
RI = NR *10 +  UC*15 +  (NC-UC)* (- 2)     
\end{equation*}
\begin{equation}
\label{equ-ri}
= 10*NR + 17*UC - 2*NC
\end{equation}

\begin{table}[]
\centering
\begin{tabular}{|l|r|r|r|r|r|r|r|}
\hline
\textbf{Developer} & \textbf{NR} & \textbf{NC}  & \textbf{UC}  & \textbf{CUD} & \textbf{ID} & \textbf{RE}&  \textbf{RI} \\
\hline
A       & 26 & 30 & 20 & 0.66 & 0.77    & 6.79 & 540 \\ 
B       & 25 & \textbf{40} & \textbf{22} & 0.55  &  0.88      & 6.72 & \textbf{544}      \\
C       & 10 & 18 & 16 & 0.89 & 1.6 & 8.61 &    336\\     
D       & 12 & 25 & 21 & 0.84 & 1.75      & \textbf{9.58} & 427\\             
E        & 1 & 5 & 5 & \textbf{1.0} & \textbf{5.0}    & 6.0 &  85  \\
F        & \textbf{30} & 5 & 4 & 0.9 &  0.13        & 5.1 & 373\\             

\hline
\end{tabular}
\caption{Review profiles of six hypothetical developers to illustrate our metrics. The highest value for each metric is highlighted using a bold text. }
\label{tab:metrics_example}
\end{table}

Table~\ref{tab:metrics_example} illustrates the metrics defined for this study based on review profiles of six hypothetical developers. In this example, Developer B is the most impactful reviewer with 25 review participation and 22 useful comments. Although, developer A participated in more code reviews,  had both higher CUD and RE scores than B, his/her RI score is lower than B due to lower number of useful comments. Although developer F had the highest review participation, his/her RI score is lower than both A and B due to his/her lowest number  of useful comments.
Developer D is the most effective reviewer with 12 review participation and 21 useful comments. Although E had the highest ID  and the highest CUD scores, his/her lack of participation in multiple reviews lowered his/her RE scores.

These illustrative examples as well as our discussions with SRBD managers suggest that this set of metrics adequately measures SRBD's  progress towards achieving more review participation from developers, promoting more useful code reviews, and encouraging developers to be more careful during reviews. Moreover, these metrics also satisfy our three requirements for the code review effectiveness metrics.

\section{Step 3: Building a framework to measure the code review effectiveness metrics}
\label{sec:measure}

\begin{figure}
    \centering
    \includegraphics[width=\linewidth]{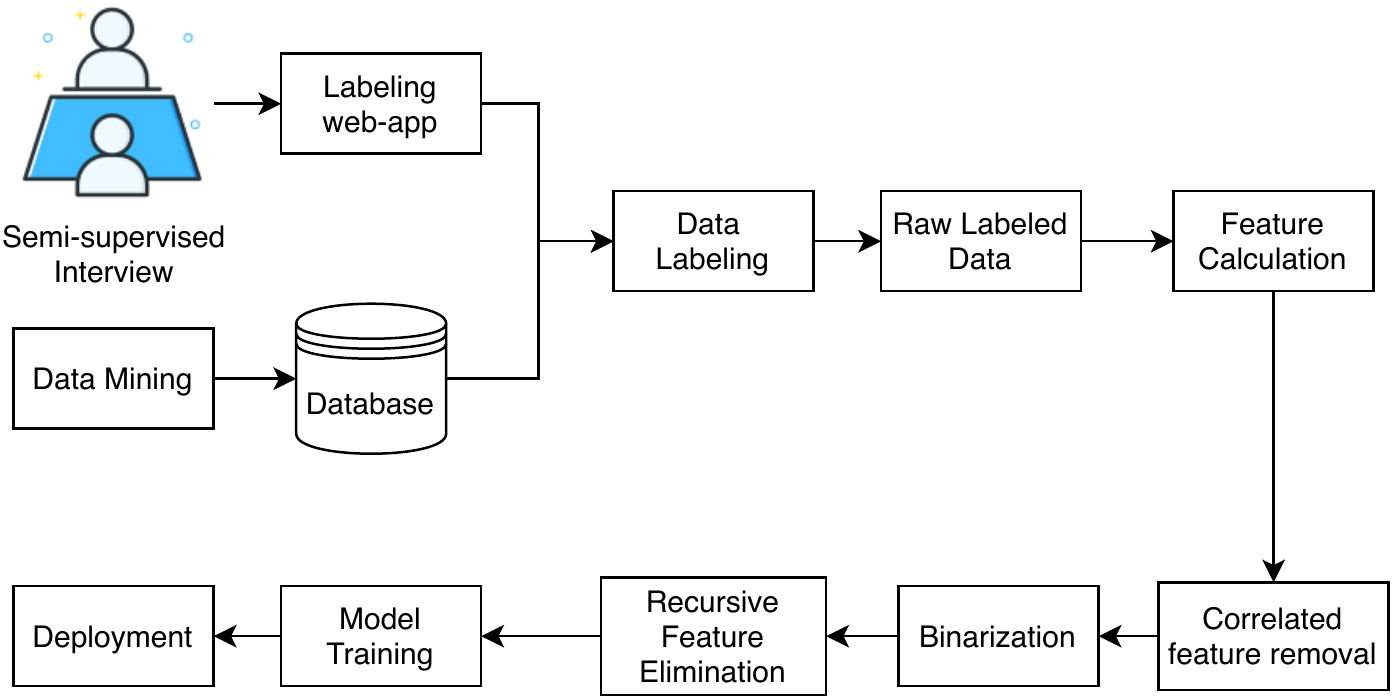}
    \caption{Research method to build an automated model to identify the usefulness of a review comment}
    \label{fig:progress-pipeline}
\end{figure}
Following step 3 of the BSC framework, this step develops a framework to measure the metrics defined in the  Section~\ref{subsec:ranking}. Those metrics require classification of each code review comment as either `useful' or `not useful'. Since manual classifications are time-consuming, we developed an automated classifier to identify the usefulness of a review comment. Figure~\ref{fig:progress-pipeline} shows an overview of the research methodology to build and evaluate our classifier. One of the researchers of this study had participated in a prior study, where he and his colleagues developed a classifier to identify useful code reviews at Microsoft~\cite{bosu}. We replicated their research methodology to build a classifier for SRBD.
In the following subsections, we describe our replication protocol, research method, and evaluation of  the classifier models

\subsection{Replication Protocol}
Replications are crucial in the software engineering domain to build knowledge through a family of experiments~\cite{basili1999building}. Shull \textit{et} al.~\cite{shull2008role} classified SE replications into two categories: 1) \textit{exact replications}, in which study procedures  are closely followed;  2) \textit{conceptual replications}, in which different research methodologies are used  to study the same set of questions. Exact replications can be further divided into two subcategories: i) \textit{dependent} and ii) \textit{independent}. In a \textit{dependent replication},  all the variables and conditions are kept as close to the original studies as possible. However, some of the aspects of the original study can be modified in an \textit{independent replication}  to fit a new context.

Since the code review tool used by SRBD is different from the one used by Microsoft as reported in Bosu \textit{et} al.'s study (referred as Microsoft study hereinafter), some of the variables can't be computed in this study. Therefore, We decided to conduct an exact independent replication~\cite{shull2008role}.
The Microsoft study was conducted using the following  three stages.

\begin{itemize}
    \item \emph{Stage 1: What makes a comment useful?} In the first stage, they conducted semi-structured interviews of seven Microsoft developers. During those interviews each developer answered questions regarding his/her perceptions of useful code reviews and labeled 20-25 code review feedback given by others to his/her  code.
    \item \emph{Stage 2: Automated classification of comment usefulness.} In the second stage, the authors of the Microsoft study manually classified 844  code review comments based on the insights gained from the first stage. Using a combined dataset of 989 code review comments (i.e., 145 classified by developers and 844 classified by the researchers), they developed a supervised learning based classifier using the Decision tree algorithm to predict the usefulness of a code review comment.  Their classifier model, which was evaluated using  10-Fold cross validations repeated 100 times, achieved a mean precision of 89.1\%, mean recall of 85.1\%, and mean accuracy of 83.4\%.
    
    \item \emph{Stage 3: Empirical study of factors influencing comment usefulness}. Using the classifier developed in stage 2, Bosu \textit{et} al. classified approximately 1.5 million code review comments from five Microsoft projects. Using this large scale dataset, they conducted an empirical study to understand the influences of various human and contextual factors on the usefulness of code reviews.
    
\end{itemize}

Rahman et al. also conducted a conceptual replication of the Microsoft study in a different commercial organization~\cite{rahman}, where they used both textual features and developer experiences to predict useful comments. While the Microsoft study focused on predicting usefulness of a comment post-completion of a review, Rahman \textit{et} al. focused on identifying potentially useful comments as soon as one is made using only pre-completion attributes of a review. Since our goal in this step is to develop an automated classifier to identify useful code review comments, post-completion, a replication of the first two stages of the Microsoft study is appropriate for our purpose. Following subsections detail our replication of the first two stages of the Microsoft study.

\subsection{Stage 1: Understanding what makes a comment useful to SRBD developers}  
Following subsections  detail our data collection methodology, interview protocol, and insights obtained from the interviews.

\subsubsection{Data collection}
We developed a Java application to mine the code reviews managed by  a Gerrit\cite{GerritCo16:online}, the code review management system of SRBD. Our miner takes a start date, an end date, and a login credential of a user as inputs. Connecting to the REST API of  SRBD's Gerrit installation,  our Gerritminer mines all the code reviews within the given date range that the login credential holder has access to. 
For this study, we were given access to seven projects that are currently under active development at SRBD. We mined a total of 9,477 code reviews, spanning four months from those seven projects.  A total of 301 distinct developers participated in those code reviews. Table \ref{tab:mined-data} shows project-wise distributions of the mined code reviews.

\begin{table}[]
\begin{tabular}{|l|l|l|l|l|l|}
\hline
\textbf{Projects} & \textbf{Changes} & {\textbf{Inline Comments}} & \textbf{Comment Labeled} &  \textbf{Usefulness \%} \\ \hline
A	& 1666	& 2930	& 650 & 81.01	\\ \hline
B	& 174	& 2010	& 639 & 81.71	\\ \hline
C	& 48	& 1871	& 607 & 73.20	\\ \hline
D	& 4937	& 1759	& 30 &  93.10 \\ \hline
E	& 1898	& 1247	& 211 & 95.67	\\ \hline
F	& 446	& 393	& 54 & 98.18	\\ \hline
G	& 308	& 183	& 13 & 53.85	\\ \hline
\textbf{Total}	& \textbf{9477}	& \textbf{10393}	& \textbf{2204} & \textbf{81.04}	\\ \hline
\end{tabular}
\caption{Project-wise Distribution of our mined dataset. Project names are anonymized due to SRBD policies. Usefulness \% indicates the ratio  of useful code review comments from that project  based on  manual labeling.}
\label{tab:mined-data}
\end{table}

\subsubsection{Developer interviews}
\label{subsec:interview}
We conducted semi-structured interviews of 14 SRBD developers from the seven selected projects to understand their perceptions of useful code reviews. 
 We carefully selected the interviewees to represent various levels of tenure years at SRBD from a pool of developers who have received at least 50 code reviews in the previous 6 months. We tried to make sure our interview responses represent different levels of experience in development and code review practice.  Table~\ref{tab:interviewee-demographics} provides an overview of the interviewees' demographics in terms of:  i) software development experience, ii) code review experience, and iii) average hours per week spent in code reviews. The duration of each interview was approximately  50 minutes, which is longer than the interview duration in the Microsoft study (i.e., 30 minutes). During our preparation for each interview, we randomly selected 50 review comments given to the code changes authored by the  interviewee.  

\begin{table}
    \centering
    \resizebox{\textwidth}{!}{
    \begin{tabular}{|l|l|r|}
    \hline
       \textbf{Demographics}  &  \textbf{Category description} & \multicolumn{1}{c|}{ \textbf{Number of developers}} \\
       \hline
     \multirow{4}{4cm}{Software development experience} & Less than one year        &  3 \\
     & Between one to three years          &   5  \\
     & Between four to six years        &   3 \\
     & More than six years        &   3 \\
     \hline

     \multirow{3}{4cm}{Tool-based code review experience} & Less than one year        &  4 \\
     & Between one to three years          &   5  \\
     & More than three years        &   5 \\
     \hline
     
     \multirow{4}{4cm}{Average hours per week spent in code reviews} & Less than five hours       &  6 \\
     & Between six to ten hours          &   5  \\
          & More than ten hours        &   3 \\
     \hline
     
    \end{tabular}
    }
    \caption{Demographics of the interviewees}
    \label{tab:interviewee-demographics}
\end{table}

Similar to the Microsoft study, our interviews were conducted in three phases. In the first phase, (about 5 minutes), we asked the interviewees demographic questions such as: job role, education, and software development experience, as well as questions to understand their perceptions of useful code reviews. In the second phase (approximately 40 minutes), we showed each interviewee our pre-selected 50 review comments for code changes that he/she has authored recently. For each of these comments, we ask an interviewee to perform the following four tasks.

\begin{enumerate}
\item Rate  whether he/she found a comment as \textit{`Useful'} or \textit{`Not Useful'}.
\item Briefly explain the rationale behind his/her classification.
 \item Mention if he/she made code changes to address that comment.
\item Classify that comment into one of our 18 predefined  categories (Figure \ref{fig:category}). This categorization scheme is based on the scheme developed by Mantyla et el.~\cite{mantyla} and later enhanced by Bosu et al.\cite{bosu} in the Microsoft study. Similar to the Microsoft study, we also provided each interviewee with a printed copy of the comment classification scheme that includes a brief description of each category to assist the interviewee's classification.
\end{enumerate}

In the final phase of the interviews (approximately 5 minutes), we asked the interviewees  to rank the comment categories based on how useful they perceive each category as. 
The interviewer took notes during the interview as well as wrote down selected ratings, categories, and explanations for  each review comment on a printed interview form.  We also recorded audios of the interviews with the interviewees’ consent. Immediately after each interview, we completed our notes with further details based on our discussions. We also listened to the audio recordings to verify the answers and notes.  We stopped conducting further interviews, when we reached an information saturation (i.e., we were no longer obtaining any new insights from interviews), after 14 interviews. During those interviews, our interviewees labeled a total 410 review comments as useful or not useful and also classified those comments into one of the 18 comment categories.

\subsubsection{Insights Gained from the Interviews:}
\label{subsec:insightfrominterview}

\begin{figure*}[t]
\pgfplotsset{width=\linewidth, height=7.5cm} 
\addtolength{\leftskip} {-.8cm}
\scalebox{1.1}{
\begin{tikzpicture}[
  every axis/.style={ 
    ybar stacked,
    ymin=0,ymax=30, 
    x tick label style={rotate=45,anchor=east},
    yticklabel=\pgfmathprintnumber{\tick}\,$\%$,
    symbolic x coords={
       Alternate Output,Design Discussion,Documentation,False Positive,Interface,Larger Defect,Logical,Naming Convention,Organization of Code,Praise,Question, Resource,Solution Approach,Support,Timing,Validation,Visual Representation,Others
    },
    xtick=data,
  bar width=12pt
  },
]
\begin{axis}[
    legend style={at={(.5,1.1)},
      anchor=south,legend columns=-1},
]
\addplot+[fill=blue!50!gray] coordinates
{(Alternate Output,2) (Design Discussion,3) (Documentation,17) (False Positive,0) (Interface,1) (Larger Defect,1) (Logical,6) (Naming Convention,12) (Organization of Code,13) (Praise,0) (Question,3) (Resource,2) (Solution Approach,3) (Support,0) (Timing,0) (Validation,2) (Visual Representation,11) (Others,3)};

\addplot+[fill=red!50!gray] coordinates
{(Alternate Output,1) (Design Discussion,0) (Documentation,3) (False Positive,6) (Interface,0) (Larger Defect,0) (Logical,0) (Naming Convention,0) (Organization of Code,2) (Praise,0) (Question,2) (Resource,1) (Solution Approach,1) (Support,0) (Timing,0) (Validation,1) (Visual Representation,5) (Others,1)};
\legend{\strut Useful, \strut Not Useful}
\end{axis}


\end{tikzpicture}
}
\caption{Distribution of various categories of code review comments as classified by SRBD developers} 
\label{fig:category}
\end{figure*}
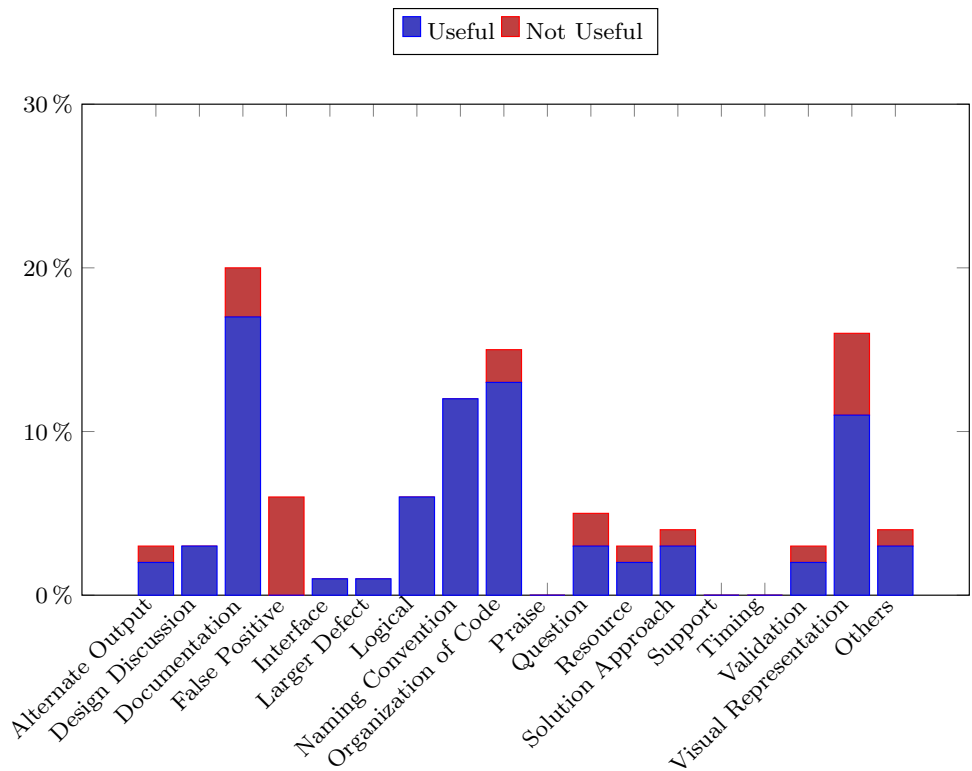

The distribution of the ratings in various review comments categorized by the authors during the interview and the labeling process (Section \ref{subsec:labeling-app}) is shown in Figure \ref{fig:category}. We  see that a much smaller number of review comments in our findings fall in the `Other' category than the previous study \cite{bosu}. This reduction is indicative that the new categories included in this labeling step (i.e., Alternate Output, Design Discussion, Praise, and Question) cover all code review comments more adequately than the 9 and 13 categories defined in previous studies respectively \cite{bosu,mantyla}. Although, expectations from  code reviews are identifying  functional defects (i.e., validation, logical, defect, resource, Interface, and support)~\cite{bacchelli2013expectations}, only 13\% code review comments belonged to those categories according to our interviewees. This result is consistent with prior studies that found less than 15\% code review comments identifying functional defects~\cite{beller2014modern,bosu,czerwonka2015code}.

According to SRBD Developers, a code review comment is   \textit{Useful} if it does at least one of the following:
\begin{enumerate}[label=(\roman*)]
\itemsep0pt
    \item Identifies a defect in code.
    \item Points out missing input validations.
    \item Helps to make code efficient and optimized.
    \item Suggests code readability improvements.
    \item Identifies logical mistakes.
    \item Identifies redundant code.
    \item Points out corner cases.
    \item Helps with code integration.
    \item Helps to detect the use of deprecated functions that will be harmful in the future merging.
    \item Suggests design improvements.
    \item Helps maintain coding standards.
    
\end{enumerate}

On the other hand a \textit{Not useful} comment belongs to one of the following:

\begin{enumerate}[label=(\roman*)]
\itemsep0pt
    \item Visual representation issues that can be also identified using static analysis tools.
    \item False positive.
    \item Code misinterpretation. 
    \item Discussion on an already resolved issue.
    \item Solution approach that the author does not agree with.
\end{enumerate}

For the following categories of comments, SRBD developers had mixed opinions as some of our interviewees considered those as \textit{Useful} while the others had a reverse opinion.  
\begin{enumerate}[label=(\roman*)]
\itemsep0pt
    \item Praise (Useful: 1, Not Useful: 4)\footnote{The numbers represent the number of interviewees that consider this type of comment as \textit{Useful} or \textit{Not Useful}}
    \item Questions asking clarifications (Useful: 1, Not Useful: 2)
    \item Suggestions to improve documentation ( Useful: 3,  Not Useful: 2) 
    
\end{enumerate}

\subsection{Stage 2: Building a classifier to automatically identify useful comments}  
Following subsections describe our approach to  create a ground truth dataset,  and training and evaluation of an automated classifier to predict the usefulness of code review comments using that dataset.

\subsubsection{Development of a ground truth dataset}

To develop a reliable supervised learning based classifier, we need a labeled training dataset (aka oracle). Although we got 410 comments classified by authors during the interviews, this dataset may not be adequate to build a reliable classifier. In the Microsoft study~\cite{bosu} as well as in its replication by Rahman et al.~\cite{rahman}, the researchers labeled additional code review comments themselves based on insights obtained from the interviews. However, code review comments labeled by the target author would be more accurate, since an author has the best knowledge of whether a comment was useful to him/her~\cite{bosu}. Therefore, we decided to get all the comments for our oracle labeled by target code authors.

\begin{figure}
  \includegraphics[width=\linewidth]{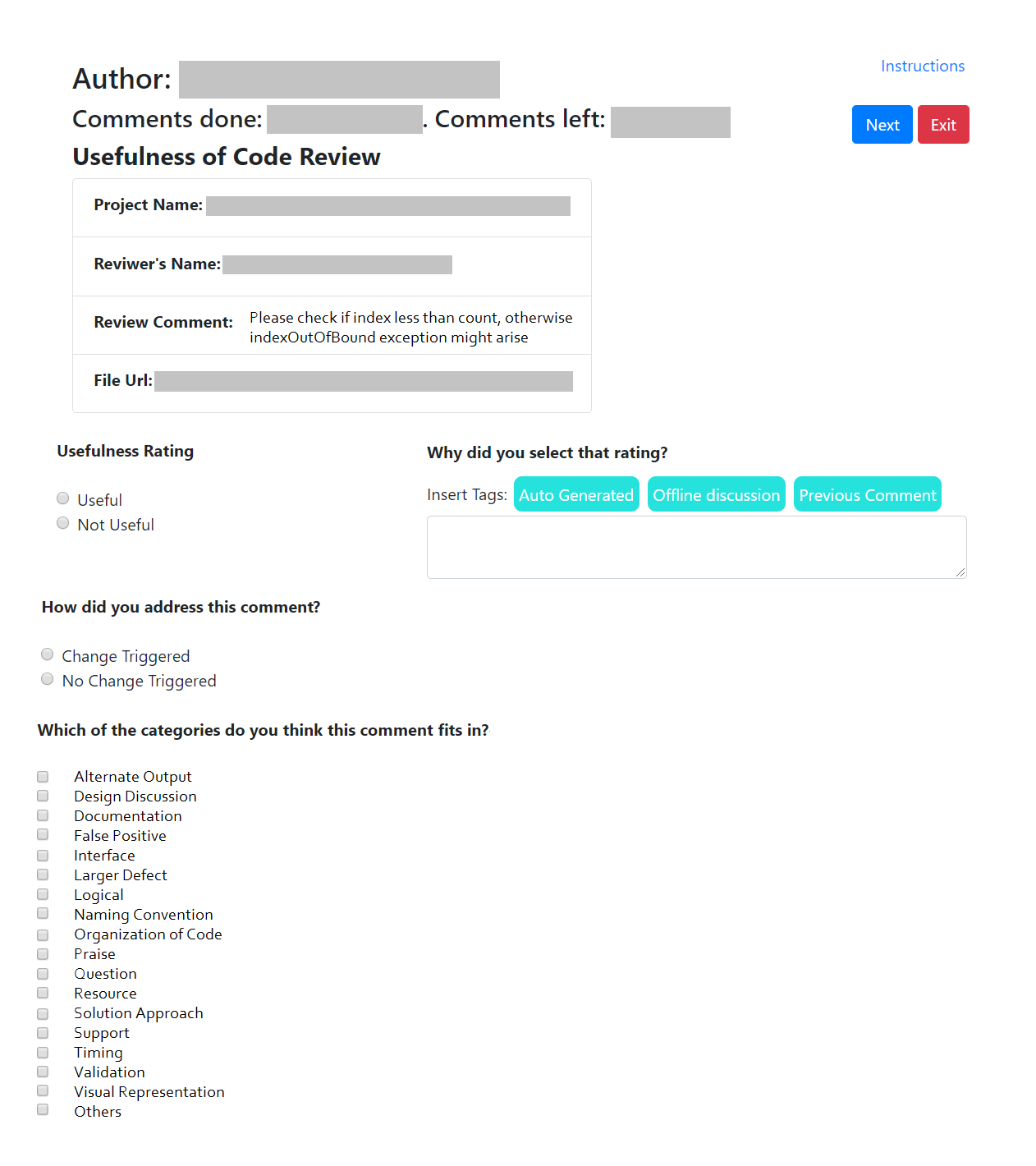}
  \caption{Screenshot of the web application to label the usefulness of a code review comment}
  \label{fig:web application}
\end{figure}

As manual data collection from the code authors is tedious, and time-consuming, we developed a web application using Python and Django framework to facilitate  asynchronous labeling of code review comments. Figure~\ref{fig:web application} shows the primary web form of our labeling application. Once authenticated, this application shows a developer  one review comment at a time, that was given to his/her code. This form also includes a link to  view the comment in Gerrit, in case the author wants to examine more context. We designed this web application to replicate our 3 phase face-to-face interview process discussed in Section \ref{subsec:interview}. A developer could login to this labeling app with his/her SRBD credentials.

We invited only those developers who  had received at least 50 code review comments during the past four months. Participation in this labeling task was voluntary and a total 40 developers signed up for this task. We organized a tutorial, where we  demonstrated the labeling app and described labeling instructions to the participants. During our demonstration,  we also  received feedback  from the participants to improve the labeling app and several of those changes were incorporated before the initiation of the labeling process. In total, 35 out of the 40 developers who signed up for this task completed the labeling process.

For each comment, an author was asked not only to rate its usefulness but also to categorize it based on the classification scheme used in our developer interviews (Section~\ref{subsec:interview}). With the help of this web application, the raters labeled a total of 1594 comments. By merging this dataset of 1594 comments with the dataset of 410 comments, labeled during face-to-face interviews, we prepared a labeled dataset of 2,004 comments.
In our dataset, 81.04\% of the data was labeled as `useful' by the code authors. 
Table \ref{tab:mined-data} demonstrates project-wise usefulness distribution.


\label{subsec:labeling-app}

\subsubsection{Feature Selection}
\label{sec:features}

Our goal  is to build a post-review completion classifier, as we do not need to build a classifier to predict comment usefulness at the time a comment is written. We considered all the features used in the Microsoft study~\cite{bosu} except the `thread status' attribute. Since SRBD's Gerrit installation did not have any feature to indicate the status of a review comment, we were forced to exclude the this attribute.  

We also included the majority of the features used in the Rahman et al.'s study~\cite{rahman}. For each of the  26 features selected for this study, Table~\ref{tab:all-features} provides a brief description, a brief rational indicating why that feature may be useful in predicting useful comments, and whether that feature was used either in the Microsoft study or in the Rahman et al.'s study. We group those features into three categories:
\begin{itemize}
\item \textit{Textual features} are computed from the code review comment text.
\item \textit{Review context features} are based on the code review process where a comment occurred.
\item \textit{Experience features} are based on the experience of the reviewer, who wrote the review. 
\end{itemize}

\begin{landscape}

\begin{table}[]

 \resizebox{19.6cm}{!}{
\begin{tabular}{|p{3.2cm}|p{8.5cm}|p{11cm}|C{1.7cm}|C{1.7cm}| }
\hline
\textbf{Feature} & \textbf{Description} & \textbf{Rationale} & \textbf{Microsoft study} & \textbf{Rahman \textit{et} al.} \\ \hline 
\hline

\multicolumn{5}{|c|}{\textbf{Textual Features}} \\ \hline

Message text & The message text is turned into a fixed dimensional vector using TF-IDF \cite{tf-idf}. & Allows the classifier to identify hidden features from the text. & \checkmark &   \\ \hline

comment\hspace{0pt}\_sentiment & The sentiment of the comment is calculated using SentiCR \cite{Ahmed-et-al-SentiCR}. & Microsoft study~\cite{bosu} found sentiment of comment as an useful feature & \checkmark & \checkmark \\ \hline

question\_ratio & Ratio of interrogative sentences in a comment. & Similar to the Microsoft developers~\cite{bosu}, most of the SRBD developers do not consider questions to be useful & & \checkmark \\ \hline

code\_element\_number & Number of source code tokens in a comment. & Inclusion of source code tokens often indicate either concerns regarding a specific section of code or specific improvement suggestion. & \checkmark  & \\ \hline

code\_element\_ratio & The ratio of source code tokens in the comment. & This is to represent the relative significance of code elements in the comment. &  & \checkmark \\ \hline 

similarity & Cosine similarity between the source code and review comment using TF-IDF \cite{tf-idf}. & Useful reviews often contain tokens from the code and are similar to the code. & & \checkmark \\ \hline 

readability & Flesch-Kincaid readability score \cite{flesch-kincaid}. & Comments that are  difficult to comprehend may cause confusions during a review~\cite{ebert2019confusion}. & \checkmark & \\ \hline

word\_count & Number of words in a comment. & A comment that is too short may be difficult comprehend. & &  \\ \hline 

stop\_word\_ratio & The ratio of stop words in the comment \cite{rahman}. & Higher ratio stop-words may indicate less informative comments with little or no semantic.  &   & \checkmark \\ \hline 
\hline

\multicolumn{5}{|c|}{\textbf{Review Context}} \\ \hline 
author\_responded & If the author responded to the review comment. & A response from the author may indicate either clarification or explicit acknowledgement of the identified issue. On the other hand, a `no response' often indicates implicit acknowledgement~\cite{bosu}. & & \\ \hline

Review interval & The time interval between the code upload and comment submission. & A hasty review may be made without properly understanding the code context~\cite{mcintosh2014impact}. &  &  \\ \hline

patch\_id & The patch number of the source code where the comment is submitted. & Comments made during the initial patches are more likely to identify major concerns, while later patches are more likely to include minor issues. & \checkmark & \\ \hline \hline

num\_patches & The total number of patchsets (i.e., review iteration) for this code review. & A larger number of patchsets may indicate, the author iteratively improving the change based on suggestions from reviewers. . & \checkmark & \\ \hline 

change\_trigger &  If the code review comment triggered a change in one of the subsequent patches. & If a comment is useful, the author is more likely to change the code to address the comment~\cite{bosu}. &  \checkmark & \\ \hline 

line\_change & The line distance of the code change if it occurred. & Distance of the change trigger might indicate if the particular comment triggered the change. & \checkmark &  \\ \hline 

confirmatory\_response & If the code author responds with \textit{``Done"}, \textit{``Fixed"}, \textit{``Removed"}, etc. & Confirmatory responses often indicate explicit acknowledgement of identified issues, and therefore an useful comment. & & \\ \hline 

gratitude & If the code author responds with \textit{``Thank you"}, \textit{``Thanks"} etc. & Gratitude comments are  often indications of explicit acknowledgement of assistance from the reviewer. & &\\ \hline 

reply\_sentiment & SentiCR \cite{Ahmed-et-al-SentiCR} sentiment score for the reply text. & Positive sentiment in the author's respond may indicate an agreement with the reviewer ( i.e., an useful comment) and vice versa. & &  \\ \hline 

is\_last\_patch & If the patch associated with the comment is the last patchset for the review. & Comments made at the last patches usually do not trigger changes and therefore are less likely to be useful unless that comment forced the author to abandon the code change.  & \checkmark & \\ \hline 

thread\_length & Total number of comments in the comment thread. & Higher number of comments in a review thread may indicate a discussion and therefore more likely to be useful. & \checkmark &  \\ \hline 

num\_participant & Number of participants in the comment thread. & More than two persons in a comment thread may indicate a concern that needs input from multiple persons. & \checkmark & \\ \hline 

review\_status & Whether the code change was `merged' or `abandoned'. & Final status of the change is important to identify whether comment made during the last patch of a review(see is\_last\_patch attribute) may be useful or not. & \checkmark & \\ \hline 
\hline

\multicolumn{5}{|c|}{\textbf{Experience Features}} \\ \hline


code\_reviewership  & The number of prior code changes of the current file the reviewer has reviewed before. & Indicates reviewer's familiarity of the current file. &  & \checkmark \\ \hline 


code\_ownership & The number of code changes the reviewer has committed for the current file. & Indicates the reviewer's familiarity with the current file. & & \checkmark \\ \hline 


reviewing\_experience & The number of code changes the reviewer has reviewed for current project. & With more review experience, a developer may get better at identifying issues. &  & \checkmark \\ \hline 

developer\_experience & The number of code changes the code author has committed for current project. & Indicates the developer's experience with the project's codebase. Developers from different experience level might perceive usefulness differently. &  & \checkmark \\ \hline 

\end{tabular}

}
\caption{Features computed for each code review comment to predict its usefulness}
\label{tab:all-features}
\end{table}

\end{landscape}

\subsubsection{Feature Computation} 
\label{feature-selection}
While the majority of the attributes selected for classifier can be directly mined from Gerrit, some of the attributes required additional computation. 
Using Gerrit's REST API and the Java diff utils library~\cite{java-diff-utils}, we implemented a Java application to automatically identify whether a review comment triggered a change within its close proximity (within 5 lines above or below). To estimate the sentiment expressed in a comment, the Microsoft study used the MSR-Splat tool. However, we replaced it with SentiCR~\cite{Ahmed-et-al-SentiCR}, an SE domain specific sentiment analysis tool, since recent research found SE domain specific tools outperforming off-the-shelf sentiment analysis tools~\cite{novielli2018benchmark}. We chose SentiCR over other tools (e.g., Senti4SD~\cite{calefato2018sentiment} and SentiStrength-SE~\cite{islam2017leveraging}), since it is customized specifically for code review texts. We used the TextStat \cite{textstat} library for calculating the Flesch-Kincaid readability score \cite{flesch-kincaid}. We curated a list of commonly occurring programming keywords (e.g. ``if ", ``class", ``switch", ``void", and ``null"), programming word patterns (e.g. camelCase, snake\_case, abc123 etc.), and reserved words in pupular programming languages (e.g. ``abstract", ``assert", ``boolean", ``break", etc.), to identify programming words in a review comment. 
To measure the similarity between a review comment and its code context, we remove coding syntaxes, stop-words from the code and comment, lemmatize  using the NLTK  \cite{nltk} library, vectorize both using TF-IDF \cite{tf-idf} and finally, calculate the cosine similarity between the code and comment vectors.

\subsubsection{Redundant Feature Elimination}

If two features are highly correlated, they are linearly dependent and have almost the same effect on the dependent variable. Therefore, dropping one of two highly correlated features reduces correlation biases~\cite{tolocsi2011classification}. We computed  a Pearson correlation matrix for our feature set and from each highly correlated ($<0.9$) feature clusters, we eliminated features having lower correlation with the ground truth usefulness score.
For example, we found that ``change\_trigger" and ``line\_change" are highly correlated with each other. Since the usefulness label's (`is\_useful')  correlation\footnote{Point biserial correlation} with ``change\_trigger" is lower than  its correlation with ``line\_change",   we eliminate  the``change\_trigger" feature here. Similarly, from the  ``num\_comments\_in\_thread", ``author\_responded" pair, we eliminated the first feature.

Since some of the features are on ratio scales (e.g., review interval, experience measures, similarity scores) , we discretized such features using {\tt qcut}, the Quantile-based discretization function from the Pandas libarary \cite{qcut}  to reduce potential overfitting. 
To identify the optimum number of features to maximize the  $F_1 score$, we use the Recursive Feature Elimination with Cross Validation (RFECV) function from the Scikit-Learn library \cite{sklearn}. RFECV eliminates the features that do not contribute significantly to a model's performance. In this stage, ``reply\_sentiment" feature is eliminated.
Details about our feature selection as well as model training steps are publicly available in Jupyter Notebook format at: \url{https://github.com/WSU-SEAL/CRA-usefulness-model}.
 

\subsubsection{Model Training and Evaluation}
\label{sec:results}

We evaluated the performances of the trained models based on three steps.
In the first step of our evaluation, we compared the performances of six classification algorithms to train our model.  In the second step,  we compared the performances of  our best performing model with human labelers from the CQA team. Finally, we compared the performance of our models against usefulness classifiers trained in prior studies~\cite{bosu,rahman}.

\underline{Models and hyperparameters: } We trained classifier models with  six supervised learning based algorithms: Decision Tree (DT) \cite{decisiontree}, Random Forest (RF) \cite{randomforest}, SVM \cite{svm}, Multi Layer Perceptron (MLPC) \cite{neuralnetwork}, XGBoost \cite{xgboost}, and Logistic Regression (LR) \cite{logisticregression}.  To find the suitable hyperparameter set for each classifier, we perform a grid search for each hyperparameter. For both RF and XGBoost, we varied with  the number of estimators (n) between 25 to 400 with an increment of 25, and  found n = 225 yielding the best results.  For the   MLPC model, we varied both the number of hidden layers and learning rates. The number of hidden layers was chosen from the set: \{ 64, 128, 256, 512\}, and the learning rate was selected from the set: \{ $1e-5$, $1e-4$, $1e-3$, $1e-2$, $1e-1$\}.  We found that the number of   hidden layers  = 256 and learning rate = $1e-4$ combination boosting the best performance.  For DT, we varied max\_depth between from 2 to 20 and found max\_depth=16 boosting the best performance. For LR, we tested with different solvers and found  the `lbfgs' solver boosting the best performance. We accepted the default parameters for the  SVM as implemented in  the LinearSVC class of the  scikit-learn library~\cite{sklearn}. We used data augmentation method SMOTE\cite{smote} to account for imbalance in our training set.

 \underline{Selection of the best model: }
Table \ref{tab:cat-results} shows the average performances of the six selected algorithms based on 20 times 10-Fold cross-validations. \boldmath$A$, \boldmath$P$, \boldmath$R$, \boldmath$F_1$, indicate the average Accuracy, average Precision, average Recall, and average $F_1$ score of the models respectively. We report the later 3 scores for both \textit{Useful} and \textit{Not Useful} classes separately.


\begin{table}[]
\centering
\begin{tabular}{|l|l|l|l|l|l|l|l|}
\hline
\multicolumn{1}{|c|}{\multirow{2}{*}{\textbf{Classifier}}} & \multicolumn{1}{c|}{\multirow{2}{*}{\textbf{$A$}}} & \multicolumn{3}{c|}{\textbf{Useful}} & \multicolumn{3}{c|}{\textbf{Not Useful}} \\ \cline{3-8} 
\multicolumn{1}{|c|}{} & \multicolumn{1}{c|}{} & \multicolumn{1}{c|}{\textbf{$P$}} & \multicolumn{1}{c|}{\textbf{$R$}} & \multicolumn{1}{c|}{\textbf{$F_1$}} & \multicolumn{1}{c|}{\textbf{$P$}} & \multicolumn{1}{c|}{\textbf{$R$}} & \multicolumn{1}{c|}{\textbf{$F_1$}} \\ \hline
       \textbf{DT}  & 81.06 & \textbf{91.02} & 85.03 & 87.89 & 49.82 & \textbf{63.91} & 55.70 \\ \hline
       \textbf{RF}  & \textbf{87.32} & 90.92 & \textbf{93.74} & \textbf{92.28} & \textbf{69.08} & 59.87 & \textbf{63.70} \\ \hline
                 \textbf{SVM}  & 63.94 & 85.70 & 67.90 & 72.30 & 30.02 & 47.93 & 30.81 \\ \hline
      \textbf{MLPC}  & 78.45 & 89.54 & 83.39 & 85.93 & 48.34 & 57.00 & 49.77 \\ \hline
             \textbf{XGBoost}  & 85.44 & 90.89 & 91.20 & 91.01 & 61.86 & 60.99 & 60.90 \\ \hline
 \textbf{LR}  & 73.77 & 87.03 & 79.49 & 83.05 & 35.87 & 49.27 & 41.25 \\ \hline
\end{tabular}
\caption{Accuracy, $Precision$, $Recall$, and $F_1$ scores for `Useful' and `Not Useful' classes for  the six models}
\label{tab:cat-results}
\end{table}

Our results (Table \ref{tab:cat-results}) suggest  that a model trained using the RF algorithm  achieves the best performances in five out of the seven measures . 
Therefore, we also compared the performances of the RF model against the models based on the other five algorithms. For this evaluation, we  obtained 200 test results for each measure (e.g., 200 precision scores for the DT models and so on) from our 20 times 10-Fold cross validations. We created the initial partition using the same $random\_state$\footnote{\url{https://scikit-learn.org/stable/modules/generated/sklearn.model_selection.KFold.html}} value to make sure that each algorithm gets the same sequence of train/test datasets. Since  the performance measures of both SVM and MLPC significantly differs from a normal distribution (Shapiro-Wilk, $p<0.05$), we used non-parametric paired sample Wilcoxon Signed-rank tests to check if the performances of the RF model significantly differ from the other five models. The results suggest (Table~\ref{tab:stat-test}) that the RF model has significantly higher accuracy,  higher recall for the `Useful' class, higher precision for the `Not Useful' class,  and higher F-scores for the both classes than all the other five models. Both DT and XGBoost have significantly higher recall for the `Not useful' class, while DT also has significantly higher precision for the `Useful' class. These results further validate that the RF model boosts the most balanced performances and outperforms the other five models. Therefore, we decided to  use RF-based models for practical applications as well as  for performance comparisons against human experts and similar prior models.

\begin{table}[]
\centering
\begin{tabular}{|l|r|r|r|r|r|r|r|}
\hline
\multicolumn{1}{|c|}{\multirow{2}{*}{\textbf{Models}}} & \multicolumn{1}{c|}{\multirow{2}{*}{\textbf{$A$}}} & \multicolumn{3}{c|}{\textbf{Useful}} & \multicolumn{3}{c|}{\textbf{Not Useful}} \\ \cline{3-8} 
\multicolumn{1}{|c|}{} & \multicolumn{1}{c|}{} & \multicolumn{1}{c|}{\textbf{$P$}} & \multicolumn{1}{c|}{\textbf{$R$}} & \multicolumn{1}{c|}{\textbf{$F_1$}} & \multicolumn{1}{c|}{\textbf{$P$}} & \multicolumn{1}{c|}{\textbf{$R$}} & \multicolumn{1}{c|}{\textbf{$F_1$}} \\ \hline

\textbf{DT} & \textbf{6.11} & \textbf{-0.28} & \textbf{8.72} & \textbf{4.31} & \textbf{18.71} & \textbf{-5.06} & \textbf{7.30} \\ \hline
\textbf{SVM} & \textbf{26.14} & \textbf{4.97} & \textbf{30.45} & \textbf{24.11} & \textbf{40.90} & \textbf{8.54} & \textbf{32.49} \\ \hline
\textbf{MLPC} & \textbf{8.76} & \textbf{1.28} & \textbf{10.28} & \textbf{6.30} & \textbf{19.80} & \textbf{1.94} & \textbf{13.08} \\ \hline
\textbf{XGBoost} & \textbf{1.96} & 0.002 & \textbf{2.70} & \textbf{1.33} & \textbf{7.69} & \textbf{-1.17} & \textbf{2.86} \\ \hline
\textbf{LR} & \textbf{13.42} & \textbf{3.65} & \textbf{14.35} & \textbf{9.18} & \textbf{32.71} & \textbf{9.45} & \textbf{21.70} \\ \hline

\end{tabular}
\caption{Performance advantages from the best performing model (i.e., Random Forest) over the other five models. Negative values indicate performance degradations. Statistically significant differences ($p<0.05$) are shown in \textbf{bold}.}
\label{tab:stat-test}
\end{table}

 \underline{Comparison against human labelers: }
As we have discussed in Section~\ref{sec:goal-definition},  two or more experienced members from the Code Quality Assessment (CQA) team used to manually evaluate the usefulness of all code reviews  to identify the best reviewer. Since our model aims to replace this manual labeling step, we compared the performances of  our best performing model against the human labelers from the CQA team. 

With this goal, we created a test dataset by randomly selecting 10\% comments from our labeled dataset and the remaining 90\% comments were assigned to the training data.
 We used the  best performing algorithm , i.e., the RF model,  to train a model based on the training dataset and evaluated its performance on the test data. We also asked an expert evaluator from the CQA team to   independently  rate the usefulness of each comment from the test data. We compute the performances of  the human labeler and our classifier against the ground truth (i.e., label received from code author).
 The top two rows of the Table \ref{tab:comparison}  show a performance comparison of our automated model against human raters. The results indicate that our automated model not only saves manual labeling efforts but also outperforms the human experts by aligning more frequently (i.e., higher accuracy and F-score) with the perceptions of the target code authors.

\begin{table}[h]
\centering
\begin{tabular}{|l|l|l|l|l|l|l|l|}
\hline
\multicolumn{1}{|c|}{\multirow{2}{*}{\textbf{Classifier}}} & \multicolumn{1}{c|}{\multirow{2}{*}{\textbf{$A$}}} & \multicolumn{3}{c|}{\textbf{Useful}} & \multicolumn{3}{c|}{\textbf{Not Useful}} \\ \cline{3-8} 
\multicolumn{1}{|c|}{} & \multicolumn{1}{c|}{} & \multicolumn{1}{c|}{\textbf{$P$}} & \multicolumn{1}{c|}{\textbf{$R$}} & \multicolumn{1}{c|}{\textbf{$F_1$}} & \multicolumn{1}{c|}{\textbf{$P$}} & \multicolumn{1}{c|}{\textbf{$R$}} & \multicolumn{1}{c|}{\textbf{$F_1$}} \\ \hline \hline
\textbf{Human Expert} & 61.88  &  86.81 &  61.72 &  72.15  &  28.99 &  \textbf{62.50}  & 39.60 \\ \hline
 \textbf{Our Classifier}  & \textbf{87.20} & \textbf{90.87} & \textbf{93.63} & \textbf{92.20} & \textbf{68.69} & 59.74 & \textbf{63.44} \\ \hline
 \hline
 \textbf{Bosu et al. \cite{bosu}} & 75.27 & 88.76 & 79.59 & 83.87 & 39.39 & 56.78 & 46.17 \\ \hline
\textbf{Rahman et al. \cite{rahman}} & 79.32 & 88.67 & 85.42 & 86.97 & 46.08 & 53.37 & 49.05 \\ \hline

\end{tabular}

\caption{Comparison  of our automated model's performances against a human rater from the CQA team, and similar models from two prior studies. Statistically significant improvements based on one-sample Wilcoxon Signed rank test ($p<0.05$) are shown in \textbf{bold}}
\label{tab:comparison}
\end{table}

\underline{Comparison against similar models: }
We compare our best performing model against similar models developed in two prior studies. While we considered all the features used by  both Bosu et al. \cite{bosu} and  Rahman et al.\cite{rahman}, we were unable to compute two attributes. First, the `Thread Status' attribute used in Bosu et al. is not available in SRBD's Gerrit. Second, the `External Library Experience' feature used in Rahman et al. could  not be computed as we were not given full access to SRBD's code repositories. Excluding those two attributes, we used the exact same algorithms and attributes to train and evaluate similar models using our dataset. 
The bottom two rows of Table \ref{tab:comparison} show the average performances of the two models based on 20 times 10-Fold crosss-validations. The results indicate that our  model outperforms both Bosu et al.'s \cite{bosu} and Rahman et al.'s \cite{rahman} models on our dataset.

\begin{figure}
    \centering
    \includegraphics[width=\linewidth]{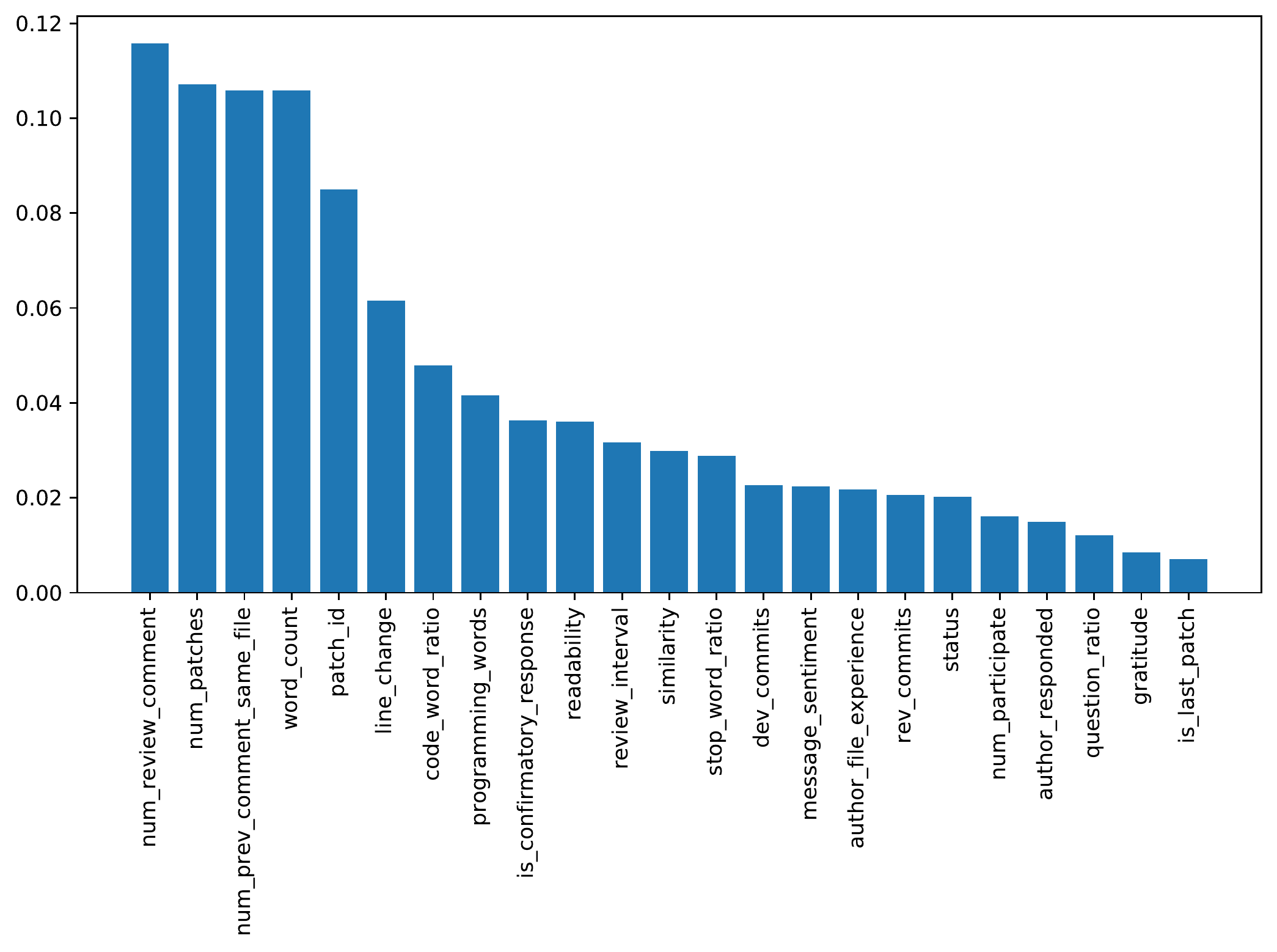}
    \caption{Relative importance of the features for the best performing model (i.e., Random Forest}
    \label{fig:feature-importance}
\end{figure}

\underline{Feature Importance Analysis:}
 We conducted a feature importance analysis for the best performing  classifier (i.e.,RF)  by calculating the Gini importance value  for each feature used in classification.  According to the results (Figure~\ref{fig:feature-importance}), review context such as number of comments, total number of patches and current patchset number,  reviewer's experience with file under review, and  textual features of the review comment had the highest influences on the fitted model.

\section{Step 4: Build a monitoring mechanism to inform the stakeholders}
\label{sec:revise}
This section implements step 4 of the BSC framework by developing Code Review Analytics (CRA), an web application that integrates the best performing model trained in the Step 3 (Section~\ref{sec:measure}). This tool enables SRBD managers and developers to continuously monitor code review effectiveness of different projects using the set of metrics defined in the Step 2 (Section ~\ref{sec:goal-metrics}) as well as individual developers. This tool allows SRBD managers to identify best reviewers during a period as well as create new initiatives to improve different projects' code reviews. Following subsections describe the development and evaluation of this CRA platform.

\begin{figure}
\centering
  \includegraphics[width=0.8\linewidth]{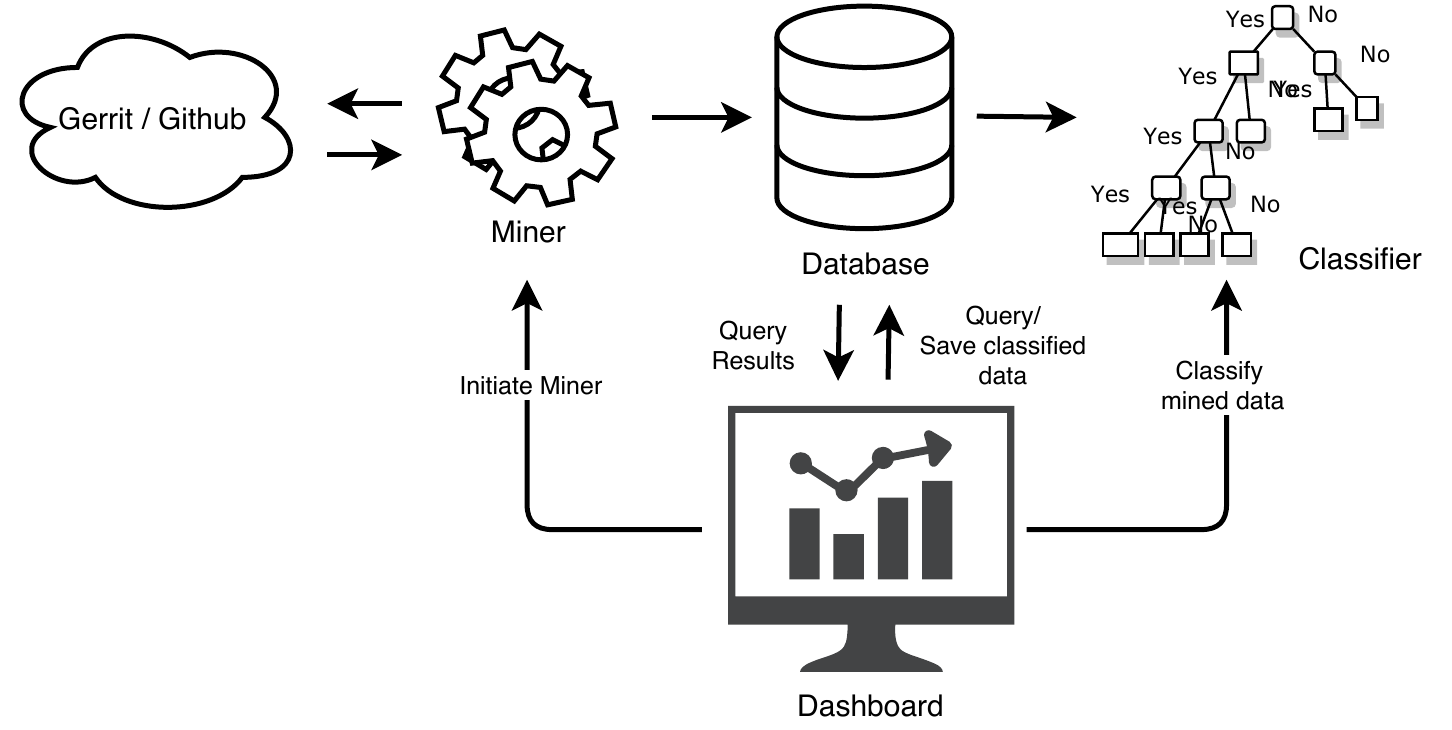}
  \caption{Workflow of the Code Review Analytics Web Application}
  \label{fig:webapp-diagram}
\end{figure}

\subsection{Development of the Code Review Analytics (CRA) tool}

\begin{figure}
  \includegraphics[width=\linewidth]{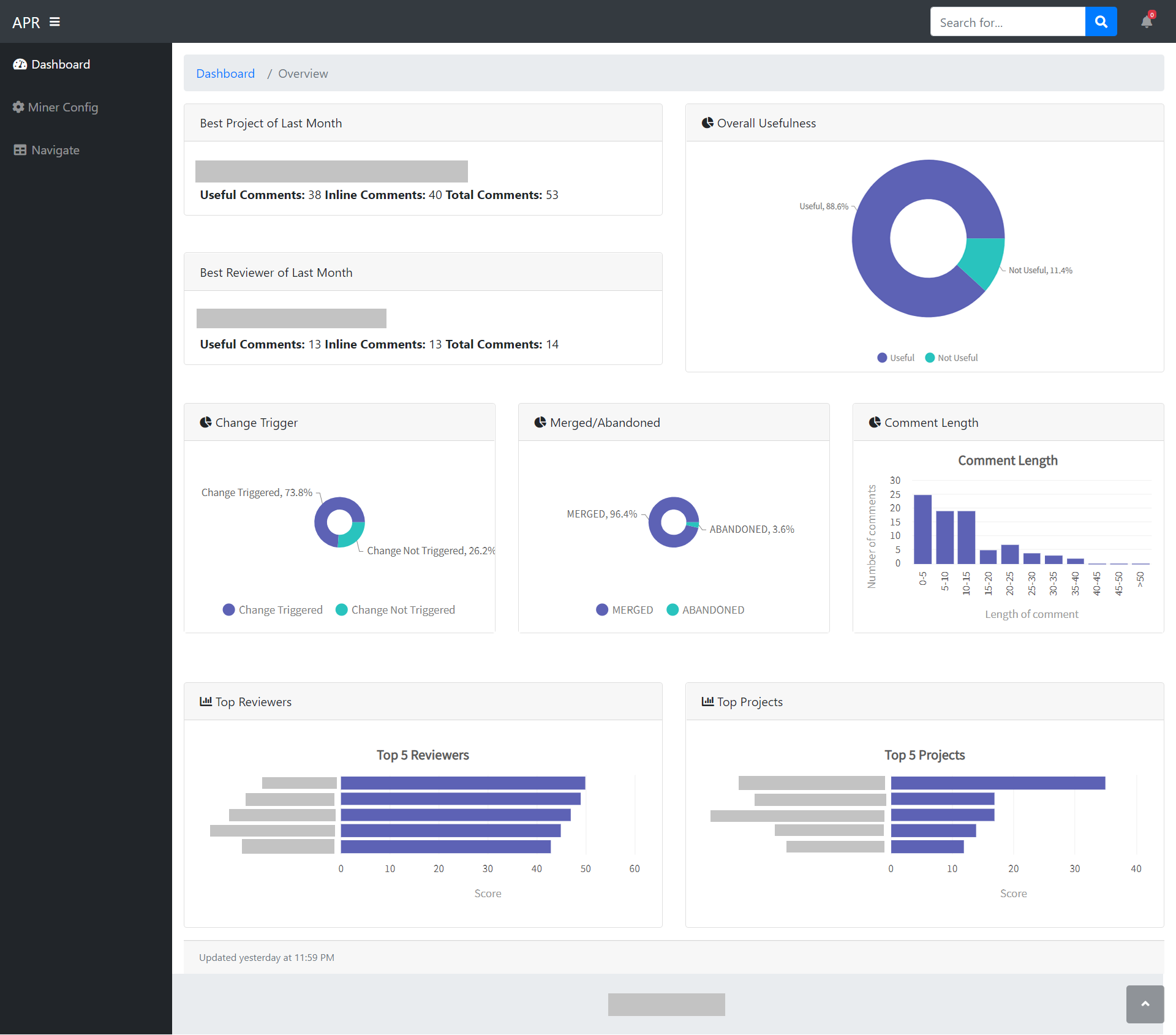}
  \caption{Screenshot of a dashboard page of the Code Review Analytics (CRA) tool }
  \label{fig:dashboard}
\end{figure}

\begin{figure}
  \includegraphics[width=\linewidth]{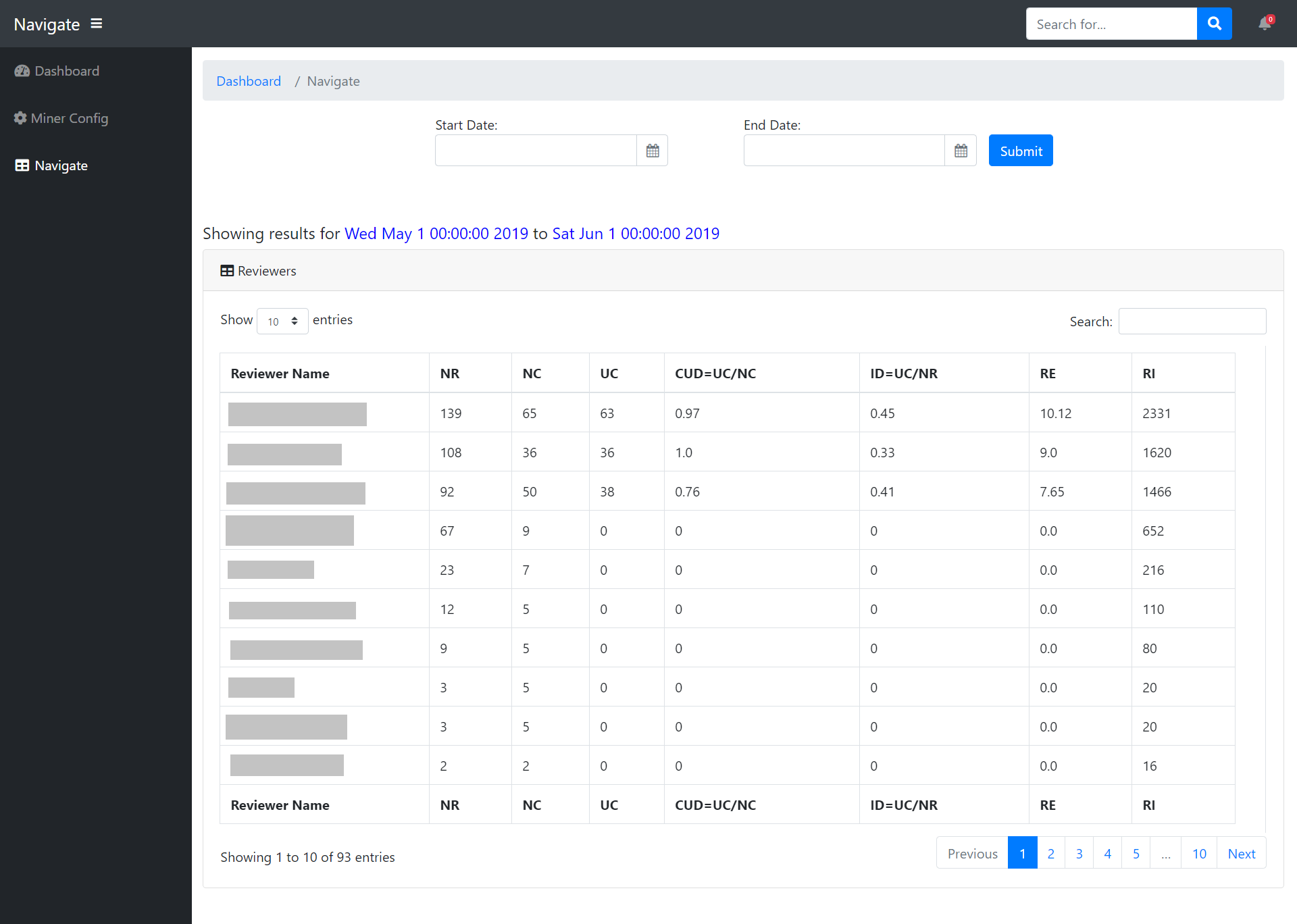} 
  \caption{ Screenshot of the Reviewer ranking page of the CRA tool showing the six measures computed for each developer}
  \label{fig:detailed-analysis}
\end{figure}

The CRA web app is built using four primary modules as illustrated in Figure~\ref{fig:webapp-diagram} and described in the following. 

\begin{enumerate}

\item \textbf{Data Miner:} The Data Miner is developed using an automated script. Upon invocation, the miner connects to a Gerrit server's REST API and downloads the code review comments created / modified since its most recent invocation. For each of the review comments, the miner computes the 16 features used in our classifier, predicts its usefulness, and stores the prediction in a database. From a configuration page, an admin user can configure the mining interval as well as manually initiate a run.

    \item \textbf{Dashboard:} After logging in to the CRA app, a user is greeted with a Dashboard page that displays a summary of code review performance in the preceding period (Fig \ref{fig:dashboard}). The best reviewer and the best project of the previous period are displayed on the top of the dashboard, along with some additional details about the overall performance of the company (e.g.,  the percent of \textit{Useful} and \textit{Not Useful} comments, the top 5 reviewers / and projects) .

\item \textbf{Reviewer \& Project Ranking Page:} Reviewer \& Project Ranking Page page allows a user to view comparison between reviewers and projects for an arbitrary date range. For reviewers and projects, the user can see their ranking based on the ranking mechanism described in Section \ref{subsec:ranking}. A snapshot of the reviewer's Detailed Analysis table at \companyname{} is shown in Fig. \ref{fig:detailed-analysis}.

\item \textbf{Reviewer or Project Details page:} Whenever the name of a reviewer or project name is clicked either from Dashboard page or Reviewer \& Project Ranking page, the user is directed to the Reviewer or Project details page, where he/she can view the performance of the reviewer or the project for each month based on the set of metrics defined in Section~\ref{sec:goal-metrics}.

\end{enumerate}

\subsection{Tool Evaluation}
After development and testing of the CRA web app, we deployed it at SRBD's internal network and ran the miner to populate the app's database with all  the code review comments authored during the last four months. We sent out invitations to a selected group of SRBD developers and managers to independently use the CRA tool for two weeks. After this initial evaluation period, we emailed a survey to nine CRA users who have voluntarily participated in the evaluation, among whom there were three managers who are involved in SRBD best reviewer selection, two members of the CQA team who have previously been involved in manual code review analysis project at SRBD, three mid-level developers, and one entry-level developer. Table~\ref{tab:tool-survey} shows the seven questions included in this user evaluation survey.

\begin{table}[]
    \centering
    \begin{tabular}{|l|p{10.5cm}|}
    \hline
    \# & \textbf{Question} \\ \hline
        Q1. &  What are your roles in SRBD?\\
        
        Q2.  &  Why do you think improving effectiveness of code reviews is important? \\
        Q3. &  How many days have you used the CRA tool during the last one week? \\
        Q4.  & On a scale 1 to 10, how useful do you find the CRA tool?\\
        Q5. & What are the insights that you are getting from the tool? \\
        Q6 & Has the tool helped you in any way in your decision making?\\
        Q7.& Do you have any suggestion to improve our tool?\\ \hline
        
    \end{tabular}
    \caption{User  evluation survey questions for the CRA application}
    \label{tab:tool-survey}
\end{table}

Majority of the invitees of our survey indicated that they had used the CRA tool multiple days during the last week. All the nine respondents rated the usefulness of the CRA tool 7 or higher on the scale of 10, with the average score being 7.44. According to the respondents, the insights they had obtained from the CRA tool include: identifying the best reviewers (3\footnote{Numbers in parentheses indicate how many CRA users of our evaluation survey mentioned this particular insight. One user may have mentioned multiple insights.}), continuous monitoring of code review performance (3), understand /improve organization's code review culture (3), identify code review effectiveness of a project or the entire organization (2), incentives for doing better reviews (1), identifying review standards (1), self-improvement as a reviewer (1). Three out of the nine respondents also indicated that the CRA tool already helped their decision making during the evaluation period. The results of our evaluation suggest that the CRA tool achieves the goal of building an automated mechanism to inform key individuals current scoreboard status and help their decision-making.

\section{Discussion} 
\label{implication}
\label{sec:implication}

Several studies related to code reviews \cite{Bosu-et-al-TSE,bosu,Kononenko-2016,rahman} foretold the promising application of automatic code review analysis at large scale. Having deployed such a tool at \companyname{}, in this section we discuss its implications.

\textbf{Self-improvement:} Continuous monitoring of reviewer performance in our analysis tool helps reviewers identify their strengths and weaknesses and  improve their weaker areas. This quicker feedback speeds up the reviewer self-improvement. Furthermore, previous studies suggest that the awareness of being observed makes individuals modify their behavior to improve productivity \cite{hawthorne}. This phenomenon is known as the \textit{Hawthorne Effect}. Knowing that the code reviews are going to be evaluated, reviewers would be more careful about reviewing code.

\textbf{Training:} The CRA tool developed in this research allows the management to observe continuous performance of individual projects and reviewers. Hence, the management can identify weaker reviewers and projects and organize necessary training sessions to the reviewers and projects in need. With the help of the analysis tool the managers can also observe the effect of their training, and update their training methodology accordingly. Visualizing the time-series of code review performance, the managers can also get an idea about how each of their decisions are affecting the code review quality, and train themselves becoming better decision-makers.

\textbf{Determine initiatives:} Our CRA tool allows the management to see the snapshot of project organization at any given time. This helps them identify weaker areas, projects, and take necessary initiatives to improve them. Our analysis tool brings the intricate organizational architecture in  interpretable graphical interface, and simplifies the decision making process.

\textbf{Motivating reviewers:} By recognizing the best reviewers and projects, an organization can  motivate all reviewers to improve their code review skill and put more focus on becoming an effective reviewer. An automated model review usefulness classification model can help identify the best reviewers with negligible manual efforts.

\textbf{Peer impression formation:} Several studies \cite{bosu-survey,Bosu-et-al-TSE,marlow2013impression} have shown that an important factor of code review is peer impression formation, that is how developers form opinions of teammates. Obtaining an accurate perception of the expertise of teammates is found to be the most important social factor of code reviews \cite{Bosu-et-al-TSE}. Our reviewer ranking tool relieves the team members from guessing each other's expertise by bringing them to broad daylight. Having each other's expertise clearly known, the team members can communicate among themselves accordingly and help the team perform optimally. 

\textbf{Usefulness model:} Using our \textit{balanced scorecard strategy (BSC)}, it is possible to build a reliable post-review usefulness classifier even under a different tool, organization. The feature-set used in our study can be directly or indirectly computed from popular code review tools. Despite being conducted on different tools, different organizations, countries, and cultures, our findings are coherent with the findings of Bosu et al. \cite{bosu}, which supports the generalisability of our approach.

\textbf{Automated program repair:} Recent study \cite{review4repair} on 14 OSS projects has shown that code reviews can be leveraged to repair defective programs using deep learning models. However, the existence of not useful reviews  can degrade performances of such models. Separation of useful reviews from non-useful reviews, using an automated model such as the one built in the research  may improve the performance of models that leverage code review information.

\section{Threats to Validity}
\label{sec:threats}
Following subsections discuss the four common categories of threats associated with an empirical study. 

\subsection{Internal validity}
Code review tools (e.g., ReviewBoard, Github pull-based reviews, and Phabricator)
could behave differently for measuring the usefulness of review comments. However, the features used in our methodology are available for all major code review tools. We think this threat is minimal for three reasons: 1) all code review tools support the same basic purpose, i.e.
detecting defects and improving the code, 2) the basic workflow (i.e., authors posting
code, reviewers commenting about code snippets, and code requiring approval from
reviewer before integration) of most of the code review tools are similar, and 3) we
did not use any Gerrit-specific feature/attribute in this study. 

The selection of participants is a source of possible bias and therefore a possible threat to internal validity. The developers selected from \companyname{} participated in at least 50 code reviews in the previous four months (either as an author or reviewer) to ensure that they can reliably label the reviews as they are used to in both reviewing and receiving reviews from peers. However, developers who participate less frequently in code reviews may have different opinions regarding useful code reviews. However, this threat may be minimal, since less than 10\% of total code reviews in SRBD during our study period belonged to such developers.

The training dataset of code review comments was curated from seven different large projects at \companyname{}. These projects are considered representative by the relevant teams due to their nature and size. Our training dataset is also representative of different programming languages, i.e. Java, Kotlin, Swift, C, C++, and Python. Although we adopted measures to minimize any biases  due to programming language, or project domain, some biases due to those factors may still exist in our models.

The ground truth dataset to train our supervised model  may become obsolete over time, and therefore degrading performance of our model. To encounter this issue, we have provided SRBD managers with the data mining and data labeling applications and detailed instructions on how to use those. SRBD developers can log in to the web app and label new code reviews. Newly labeled data can be augmented to the existing dataset to retrain the usefulness classifier.

\subsection{Construct Validity}
The primary source of construct validity is the variations among the criteria used by different developers to  consider a code review comment as `useful'. For example, during our interviews, we found that some developers were more generous than others, and  have expansive definitions of what they consider as `useful'. While we attempted to mitigate such biases by providing labelers with some guidelines through a seminar, such biases still remain in our dataset. Regardless, this threat may be minimal, since our best performing model still achieves an accuracy of $\approx$ 87\% and an F-score of $\approx$92\%.

\subsection{External Validity}
Results based on a single project or even a handful of projects, or a single organization can be subject to lack of external validity. Therefore, one may claim that  empirical research within one organization or one project is  inadequate, provides little value for the academic community, and does not contribute to scientific development.  Moreover, historical evidence provides several examples of individual cases that contributed to discovery in physics, economics, and social science (see ``Five misunderstandings about case-study research'' by Flyvjerg~\cite{flyvbjerg2006five}). Even in the SE domain case studies of a single project~\cite{di2016security,camilo2015bugs,khomh2012faster,mockus2000case} as well as single organization~\cite{Bosu-et-al-TSE,bosu,sadowski2018modern} have provided important insights. Moreover, following the study protocol of this study, one can develop similar models or tools for another organization. To promote such replications, we have made our Jupyter notebooks publicly available on Github.

\subsection{Conclusion Validity}
We adopted several measures to minimize biases during training and evaluation of our models to predict the usefulness of a review comment. First, we got each code review comment labeled by the target author, since he/she is the best person to determine whether a comment helped him/her to improve a code change. Second, we removed redundant features to reduce correlation biases. Finally, we took several initiatives to reduce potential over-fitting by:  1) discretizing features on  ratio scales, 2) using recursive feature elimination with cross validation to eliminate unnecessary features, 3) evaluating models based on 20 times 10-fold cross validations. Therefore, we do not envision any significant threats to conclusion validity for our models.

\section{Conclusion}
\label{sec:conclusion}
In this study, we developed and evaluated a solution based on the Balanced Scorecard strategy framework to monitor code review effectiveness at SRBD, a commercial software development organization. Our solution also assists SRBD managers to identify opportunities to improve the effectiveness of code reviews by providing them continuous access to various measures and reports. 
Prior to the deployment of our solution, SRBD used manual assessments by CQA members to track code review effectiveness as well as to identify the best reviewer(s) for a particular period. However, such manual assessments were: 1) prone to inconsistencies, 2) time consuming and not scalable, 3) prone to delayed assessments, and  4) failing to provide insights to pinpoint areas of concern. Our automated solution addresses these four shortcomings of SRBD's prior assessment mechanism. 

Moreover, our automated model to identify useful code reviews not only outperforms the manual assessments from human labelers but also significantly reduces such manual labeling efforts. 
Using our solution SRBD managers can identify areas of concern quickly and take  immediate necessary actions. Our solution's transparent mechanism to identify and showcase the effective reviewers can motivate useful code reviews and  help improve the code review culture at SRBD.
Our reviewer ranking dashboard also helps management to assign better reviewers, transfer reviewers to right teams, or arrange training sessions for new or under-performing reviewers. A developer can also receive training implicitly by observing his/ her review that is identified as `useful'/`not useful' with a particular context. 

We are also encouraged by the positive evaluation from the users of our solution. We believe that other organizations may be encouraged by the successful deployment and observe positive benefits from our solution at SRBD and develop similar solutions to monitor their code review effectiveness as well as identify potential improvement opportunities.


\section*{Acknowledgment}

Work conducted by Dr. Amiangshu Bosu for this research is partially supported by the US National Science Foundation under Grant No. 1850475.
Any opinions, findings, and conclusions or recommendations expressed in this material are those of the author(s) and do not necessarily reflect the views of the National Science Foundation.

Work conducted for this research is also partially supported by a research grant provided by the Samsung Research Bangladesh.

\bibliography{references}

\begin{thebibliography}{10}
\providecommand{\url}[1]{{#1}}
\providecommand{\urlprefix}{URL }
\expandafter\ifx\csname urlstyle\endcsname\relax
  \providecommand{\doi}[1]{DOI~\discretionary{}{}{}#1}\else
  \providecommand{\doi}{DOI~\discretionary{}{}{}\begingroup
  \urlstyle{rm}\Url}\fi

\bibitem{GerritCo16:online}
Gerrit code review - rest api.
\newblock
  \url{https://gerrit-review.googlesource.com/Documentation/rest-api.html}.
\newblock (Accessed on 09/27/2019)

\bibitem{nltk}
Natural language toolkit — nltk 3.5 documentation.
\newblock \url{https://www.nltk.org/}.
\newblock (Accessed on 12/06/2020)

\bibitem{qcut}
pandas.qcut — pandas 1.1.5 documentation.
\newblock
  \url{https://pandas.pydata.org/pandas-docs/stable/reference/api/pandas.qcut.html}.
\newblock (Accessed on 12/24/2020)

\bibitem{textstat}
textstat · pypi.
\newblock \url{https://pypi.org/project/textstat/}.
\newblock (Accessed on 12/06/2020)

\bibitem{Ahmed-et-al-SentiCR}
Ahmed, T., Bosu, A., Iqbal, A., Rahimi, S.: {SentiCR: A Customized Sentiment
  Analysis Tool for Code Review Interactions}.
\newblock In: 32nd IEEE/ACM International Conference on Automated Software
  Engineering (NIER track), ASE '17 (2017)

\bibitem{bacchelli2013expectations}
Bacchelli, A., Bird, C.: Expectations, outcomes, and challenges of modern code
  review.
\newblock In: Proceedings of the 2013 international conference on software
  engineering, pp. 712--721. IEEE Press (2013)

\bibitem{barnett2015helping}
Barnett, M., Bird, C., Brunet, J., Lahiri, S.K.: Helping developers help
  themselves: Automatic decomposition of code review changesets.
\newblock In: Proceedings of the 37th International Conference on Software
  Engineering-Volume 1, pp. 134--144. IEEE Press (2015)

\bibitem{basili1999building}
Basili, V.R., Shull, F., Lanubile, F.: Building knowledge through families of
  experiments.
\newblock IEEE Transactions on Software Engineering \textbf{25}(4), 456--473
  (1999)

\bibitem{beller2014modern}
Beller, M., Bacchelli, A., Zaidman, A., Juergens, E.: Modern code reviews in
  open-source projects: Which problems do they fix?
\newblock In: Proceedings of the 11th working conference on mining software
  repositories, pp. 202--211 (2014)

\bibitem{di2016security}
di~Biase, M., Bruntink, M., Bacchelli, A.: A security perspective on code
  review: The case of chromium.
\newblock In: 2016 IEEE 16th International Working Conference on Source Code
  Analysis and Manipulation (SCAM), pp. 21--30. IEEE (2016)

\bibitem{bosu-survey}
{Bosu}, A., {Carver}, J.C.: Impact of peer code review on peer impression
  formation: A survey.
\newblock In: 2013 ACM / IEEE International Symposium on Empirical Software
  Engineering and Measurement, pp. 133--142 (2013).
\newblock \doi{10.1109/ESEM.2013.23}

\bibitem{Bosu-et-al-TSE}
Bosu, A., Carver, J.C., Bird, C., Orbeck, J., Chockley, C.: Process aspects and
  social dynamics of contemporary code review: Insights from open source
  development and industrial practice at microsoft.
\newblock IEEE Transactions on Software Engineering \textbf{43}(1), 56--75
  (2017).
\newblock \doi{10.1109/TSE.2016.2576451}

\bibitem{bosu}
Bosu, A., Greiler, M., Bird, C.: Characteristics of {Useful} {Code} {Reviews}:
  {An} {Empirical} {Study} at {Microsoft}.
\newblock In: 2015 {IEEE}/{ACM} 12th {Working} {Conference} on {Mining}
  {Software} {Repositories}, pp. 146--156. IEEE, Florence, Italy (2015).
\newblock \doi{10.1109/MSR.2015.21}.
\newblock \urlprefix\url{http://ieeexplore.ieee.org/document/7180075/}

\bibitem{sklearn}
Buitinck, L., Louppe, G., Blondel, M., Pedregosa, F., Mueller, A., Grisel, O.,
  Niculae, V., Prettenhofer, P., Gramfort, A., Grobler, J., Layton, R.,
  VanderPlas, J., Joly, A., Holt, B., Varoquaux, G.: {API} design for machine
  learning software: experiences from the scikit-learn project.
\newblock In: ECML PKDD Workshop: Languages for Data Mining and Machine
  Learning, pp. 108--122 (2013)

\bibitem{calefato2018sentiment}
Calefato, F., Lanubile, F., Maiorano, F., Novielli, N.: Sentiment polarity
  detection for software development.
\newblock Empirical Software Engineering \textbf{23}(3), 1352--1382 (2018)

\bibitem{camilo2015bugs}
Camilo, F., Meneely, A., Nagappan, M.: Do bugs foreshadow vulnerabilities?: a
  study of the chromium project.
\newblock In: Proceedings of the 12th Working Conference on Mining Software
  Repositories, pp. 269--279. IEEE Press (2015)

\bibitem{smote}
Chawla, N.V., Bowyer, K.W., Hall, L.O., Kegelmeyer, W.P.: {SMOTE}: {Synthetic}
  {Minority} {Over}-sampling {Technique}.
\newblock Journal of Artificial Intelligence Research \textbf{16}, 321--357
  (2002).
\newblock \doi{10.1613/jair.953}.
\newblock \urlprefix\url{https://jair.org/index.php/jair/article/view/10302}

\bibitem{xgboost}
Chen, T., Guestrin, C.: Xgboost: A scalable tree boosting system.
\newblock In: Proceedings of the 22nd acm sigkdd international conference on
  knowledge discovery and data mining, pp. 785--794. ACM (2016)

\bibitem{9425884}
Chouchen, M., Ouni, A., Kula, R.G., Wang, D., Thongtanunam, P., Mkaouer, M.W.,
  Matsumoto, K.: Anti-patterns in modern code review: Symptoms and prevalence.
\newblock In: 2021 IEEE International Conference on Software Analysis,
  Evolution and Reengineering (SANER), pp. 531--535 (2021).
\newblock \doi{10.1109/SANER50967.2021.00060}

\bibitem{whoreview}
Chouchen, M., Ouni, A., Mkaouer, M.W., Kula, R.G., Inoue, K.: Whoreview: A
  multi-objective search-based approach for code reviewers recommendation in
  modern code review.
\newblock Applied Soft Computing \textbf{100}, 106908 (2021).
\newblock \doi{https://doi.org/10.1016/j.asoc.2020.106908}.
\newblock
  \urlprefix\url{https://www.sciencedirect.com/science/article/pii/S1568494620308462}

\bibitem{cohen2006best}
Cohen, J., Brown, E., DuRette, B., Teleki, S.: Best kept secrets of peer code
  review.
\newblock Smart Bear Somerville (2006)

\bibitem{czerwonka2015code}
Czerwonka, J., Greiler, M., Tilford, J.: Code reviews do not find bugs: how the
  current code review best practice slows us down.
\newblock In: Proceedings of the 37th International Conference on Software
  Engineering-Volume 2, pp. 27--28. IEEE Press (2015)

\bibitem{8668024}
Ebert, F., Castor, F., Novielli, N., Serebrenik, A.: Confusion in code reviews:
  Reasons, impacts, and coping strategies.
\newblock In: 2019 IEEE 26th International Conference on Software Analysis,
  Evolution and Reengineering (SANER), pp. 49--60 (2019).
\newblock \doi{10.1109/SANER.2019.8668024}

\bibitem{ebert2019confusion}
Ebert, F., Castor, F., Novielli, N., Serebrenik, A.: Confusion in code reviews:
  Reasons, impacts, and coping strategies.
\newblock In: 2019 IEEE 26th International Conference on Software Analysis,
  Evolution and Reengineering (SANER), pp. 49--60. IEEE (2019)

\bibitem{Fagan}
Fagan, M.E.: Design and code inspections to reduce errors in program
  development.
\newblock IBM Syst. J. \textbf{15}(3), 182--211 (1976).
\newblock \doi{10.1147/sj.153.0182}.
\newblock \urlprefix\url{http://dx.doi.org/10.1147/sj.153.0182}

\bibitem{flesch-kincaid}
Flesch, R.: Flesch--kincaid readability test.
\newblock Retrieved October \textbf{26}, 2007 (2007)

\bibitem{flyvbjerg2006five}
Flyvbjerg, B.: Five misunderstandings about case-study research.
\newblock Qualitative inquiry \textbf{12}(2), 219--245 (2006)

\bibitem{developersgame}
{Fracz}, W., {Dajda}, J.: Developers' game: A preliminary study concerning a
  tool for automated developers assessment.
\newblock In: 2018 IEEE International Conference on Software Maintenance and
  Evolution (ICSME), pp. 695--699 (2018).
\newblock \doi{10.1109/ICSME.2018.00079}

\bibitem{hatton2008testing}
Hatton, L.: Testing the value of checklists in code inspections.
\newblock IEEE software \textbf{25}(4), 82--88 (2008)

\bibitem{hirao2016impact}
Hirao, T., Ihara, A., Ueda, Y., Phannachitta, P., Matsumoto, K.i.: The impact
  of a low level of agreement among reviewers in a code review process.
\newblock In: IFIP International Conference on Open Source Systems, pp.
  97--110. Springer (2016)

\bibitem{randomforest}
Ho, T.K.: Random decision forests.
\newblock In: Proceedings of 3rd international conference on document analysis
  and recognition, vol.~1, pp. 278--282. IEEE (1995)

\bibitem{hofner2011fostering}
Hofner, G., Mani, V., Nambiar, R., Apte, M.: Fostering a high-performance
  culture in offshore software engineering teams using balanced scorecards and
  project scorecards.
\newblock In: 2011 IEEE Sixth International Conference on Global Software
  Engineering, pp. 35--39. IEEE (2011)

\bibitem{neuralnetwork}
Hopfield, J.J.: Artificial neural networks.
\newblock IEEE Circuits and Devices Magazine \textbf{4}(5), 3--10 (1988)

\bibitem{logisticregression}
Hosmer~Jr, D.W., Lemeshow, S., Sturdivant, R.X.: Applied logistic regression,
  vol. 398.
\newblock John Wiley \& Sons (2013)

\bibitem{review4repair}
Huq, F., Hasan, M., Pantho, M.A.H., Mahbub, S., Iqbal, A., Ahmed, T.:
  Review4repair: Code review aided automaticprogram repairing.
\newblock arXiv preprint arXiv:2010.01544  (2020)

\bibitem{islam2017leveraging}
Islam, M.R., Zibran, M.F.: Leveraging automated sentiment analysis in software
  engineering.
\newblock In: 2017 IEEE/ACM 14th International Conference on Mining Software
  Repositories (MSR), pp. 203--214. IEEE (2017)

\bibitem{decisiontree}
Jin, C., De-Lin, L., Fen-Xiang, M.: An improved id3 decision tree algorithm.
\newblock In: 2009 4th International Conference on Computer Science \&
  Education, pp. 127--130. IEEE (2009)

\bibitem{balancedscorecard}
Kaplan, R.S., Norton, D.P., et~al.: The balanced scorecard: measures that drive
  performance  (1992)

\bibitem{khomh2012faster}
Khomh, F., Dhaliwal, T., Zou, Y., Adams, B.: Do faster releases improve
  software quality?: an empirical case study of mozilla firefox.
\newblock In: Proceedings of the 9th IEEE Working Conference on Mining Software
  Repositories, pp. 179--188. IEEE Press (2012)

\bibitem{Kononenko-2016}
Kononenko, O., Baysal, O., Godfrey, M.W.: Code review quality: How developers
  see it.
\newblock In: Proceedings of the 38th International Conference on Software
  Engineering, ICSE '16, pp. 1028--1038. ACM, New York, NY, USA (2016).
\newblock \doi{10.1145/2884781.2884840}.
\newblock \urlprefix\url{http://doi.acm.org/10.1145/2884781.2884840}

\bibitem{Kononenko-emp}
{Kononenko}, O., {Baysal}, O., {Guerrouj}, L., {Cao}, Y., {Godfrey}, M.W.:
  Investigating code review quality: Do people and participation matter?
\newblock In: 2015 IEEE International Conference on Software Maintenance and
  Evolution (ICSME), pp. 111--120 (2015)

\bibitem{mair2002balanced}
Mair, S.: A balanced scorecard for a small software group.
\newblock IEEE software \textbf{19}(6), 21--27 (2002)

\bibitem{marlow2013impression}
Marlow, J., Dabbish, L., Herbsleb, J.: Impression formation in online peer
  production: activity traces and personal profiles in github.
\newblock In: Proceedings of the 2013 conference on Computer supported
  cooperative work, pp. 117--128 (2013)

\bibitem{marr2003automating}
Marr, B., Neely, A.: Automating the balanced scorecard--selection criteria to
  identify appropriate software applications.
\newblock Measuring Business Excellence  (2003)

\bibitem{hawthorne}
McCarney, R., Warner, J., Iliffe, S., Van~Haselen, R., Griffin, M., Fisher, P.:
  The hawthorne effect: a randomised, controlled trial.
\newblock BMC medical research methodology \textbf{7}(1), 30 (2007)

\bibitem{mcintosh2014impact}
McIntosh, S., Kamei, Y., Adams, B., Hassan, A.E.: The impact of code review
  coverage and code review participation on software quality: A case study of
  the qt, vtk, and itk projects.
\newblock In: Proceedings of the 11th Working Conference on Mining Software
  Repositories, pp. 192--201 (2014)

\bibitem{mantyla}
{Mäntylä}, M.V., {Lassenius}, C.: What types of defects are really discovered
  in code reviews?
\newblock IEEE Transactions on Software Engineering \textbf{35}(3), 430--448
  (2009).
\newblock \doi{10.1109/TSE.2008.71}

\bibitem{mockus2000case}
Mockus, A., Fielding, R.T., Herbsleb, J.: A case study of open source software
  development: the apache server.
\newblock In: Proceedings of the 22nd international conference on Software
  engineering, pp. 263--272. Acm (2000)

\bibitem{java-diff-utils}
Naumenko, D.: Java diff utils.
\newblock \url{https://github.com/dnaumenko/java-diff-utils} (2018)

\bibitem{novielli2018benchmark}
Novielli, N., Girardi, D., Lanubile, F.: A benchmark study on sentiment
  analysis for software engineering research.
\newblock In: 2018 IEEE/ACM 15th International Conference on Mining Software
  Repositories (MSR), pp. 364--375. IEEE (2018)

\bibitem{papalexandris2004implementing}
Papalexandris, A., Ioannou, G., Prastacos, G.P.: Implementing the balanced
  scorecard in greece: a software firm’s experience.
\newblock Long Range Planning \textbf{37}(4), 351--366 (2004)

\bibitem{rahman-correct}
Rahman, M.M., Roy, C.K., Collins, J.A.: Correct: code reviewer recommendation
  in github based on cross-project and technology experience.
\newblock In: Proceedings of the 38th International Conference on Software
  Engineering Companion, pp. 222--231 (2016)

\bibitem{rahman}
Rahman, M.M., Roy, C.K., Kula, R.G.: Predicting usefulness of code review
  comments using textual features and developer experience.
\newblock In: Proceedings of the 14th International Conference on Mining
  Software Repositories, MSR '17, p. 215–226. IEEE Press (2017)

\bibitem{rigby2013convergent}
Rigby, P.C., Bird, C.: Convergent contemporary software peer review practices.
\newblock In: Proceedings of the 2013 9th Joint Meeting on Foundations of
  Software Engineering, pp. 202--212. ACM (2013)

\bibitem{rigby2006preliminary}
Rigby, P.C., German, D.M.: A preliminary examination of code review processes
  in open source projects.
\newblock Tech. rep., Technical Report DCS-305-IR, University of Victoria
  (2006)

\bibitem{sadowski2018modern}
Sadowski, C., S{\"o}derberg, E., Church, L., Sipko, M., Bacchelli, A.: Modern
  code review: a case study at google.
\newblock In: Proceedings of the 40th International Conference on Software
  Engineering: Software Engineering in Practice, pp. 181--190. ACM (2018)

\bibitem{tf-idf}
Salton, G., Buckley, C.: Term-weighting approaches in automatic text retrieval.
\newblock Inf. Process. Manage. \textbf{24}(5), 513--523 (1988).
\newblock \doi{10.1016/0306-4573(88)90021-0}.
\newblock \urlprefix\url{http://dx.doi.org/10.1016/0306-4573(88)90021-0}

\bibitem{shull2008role}
Shull, F.J., Carver, J.C., Vegas, S., Juristo, N.: The role of replications in
  empirical software engineering.
\newblock Empirical software engineering \textbf{13}(2), 211--218 (2008)

\bibitem{so-reputation}
StackOverflow: What is reputation? how do i earn (and losek) it?
\newblock https://stackoverflow.com/help/whats-reputation

\bibitem{svm}
Suykens, J.A., Vandewalle, J.: Least squares support vector machine
  classifiers.
\newblock Neural processing letters \textbf{9}(3), 293--300 (1999)

\bibitem{thongtanunam-dynamics}
Thongtanunam, P., Hassan, A.E.: Review dynamics and their impact on software
  quality.
\newblock IEEE Transactions on Software Engineering pp. 1--1 (2020).
\newblock \doi{10.1109/TSE.2020.2964660}

\bibitem{thongtanunam2017review}
Thongtanunam, P., McIntosh, S., Hassan, A.E., Iida, H.: Review participation in
  modern code review.
\newblock Empirical Software Engineering \textbf{22}(2), 768--817 (2017)

\bibitem{thongtanunam2015should}
Thongtanunam, P., Tantithamthavorn, C., Kula, R.G., Yoshida, N., Iida, H.,
  Matsumoto, K.i.: Who should review my code? a file location-based
  code-reviewer recommendation approach for modern code review.
\newblock In: 2015 IEEE 22nd International Conference on Software Analysis,
  Evolution, and Reengineering (SANER), pp. 141--150. IEEE (2015)

\bibitem{tolocsi2011classification}
Tolo{\c{s}}i, L., Lengauer, T.: Classification with correlated features:
  unreliability of feature ranking and solutions.
\newblock Bioinformatics \textbf{27}(14), 1986--1994 (2011)

\end{thebibliography}

\end{document}